\documentclass[reqno, 11pt, a4paper]{amsart}
\usepackage[dvipsnames]{xcolor}
\usepackage{tikz-cd}
\usepackage[utf8]{inputenc}
\usepackage[T1]{fontenc}
\usepackage{amsmath}
\usepackage{amsfonts,amssymb}
\usetikzlibrary{matrix}
\usepackage{soul}

\usepackage{marginnote}
\makeatletter
\newcommand{\vast}{\bBigg@{14.030}}
\newcommand{\Vast}{\bBigg@{18.30}}
\makeatother
\newcommand{\marginttt}[1]{}
\usepackage{tabularx}

\newcommand{\sa}{_{\text{\tiny s.a.}}}

\newcommand{\rojo}[1]{}

\definecolor{azulESI}{HTML}{1266AE}

\definecolor{AZULESI}{HTML}{1266AE}
\usepackage{enumerate}

\usepackage{nicematrix}
\usepackage{soul}
\usepackage[utf8]{inputenc}

\makeatletter
\ifcase \@ptsize \relax
  \newcommand{\nano}{\@setfontsize\miniscule{3.5}{4.5}}
\or
  \newcommand{\nano}{\@setfontsize\miniscule{4.5}{5.5}}%
\or
  \newcommand{\nano}{\@setfontsize\miniscule{4.5}{5.5}}%
\fi
\makeatother

\newcommand{\balita}{\raisebox{1.9pt}{\text{\nano$\bullet$\hspace{.7pt}}}}

\usepackage{tikz}
\usetikzlibrary{automata,arrows,topaths}
\usepackage{caption}

\usepackage[bookmarks=false]{hyperref}
\hypersetup{
         colorlinks   = true,
         citecolor   =azulESI,
        urlcolor = azulESI
}
\hypersetup{linkcolor=azulESI}

\newcommand{\pS}{{\mathscr{pS}}}

\newcommand{\inv}{^{-1}}

\usepackage[scr=boondoxo]{mathalfa}

\usepackage{float}
\usepackage{subcaption}
\usepackage{amsthm}
\numberwithin{equation}{section}
\newtheoremstyle{mytheoremstyle} 
    {10pt}                    
    {8pt}                    
    {\itshape}                   
    {}                           
    {\scshape}                   
    {.}                          
    {.5em}                       
    {}  

\makeatletter
\newcommand{\leqnomode}{\tagsleft@true}
\newcommand{\reqnomode}{\tagsleft@false}
\makeatother
\theoremstyle{mytheoremstyle}

\newtheorem{theorem}{Theorem}[section]

 \newtheoremstyle{definition} 
    {8pt}                    
    {5pt}                    
    {}                   
    {}                           
    {\scshape}                   
    {.}                          
    {.5em}                       
    {}  

 \theoremstyle{definition}
 \newtheorem{definition}[theorem]{Definition}

 \newtheorem{remark}[theorem]{Remark}

\newcommand{\N}{\mathbb{Z}_{>0}}
\newcommand{\Z}{\mathbb{Z}}

\newcommand{\Znn}{\mathbb{Z}_{\geq 0}}
\newcommand{\C}{\mathbb{C}}
\newcommand{\R}{\mathbb{R}}

\newcommand{\runterhalb}[1] {\raisebox{-.1\height}{#1}}

\newcommand{\sumsub}[1]{\sum_{\substack{#1}}}

\newcommand{\dif}{{\mathrm{d}}}

\newcommand{\M}[1]{M_{#1}(\mathbb{C})}
\newcommand{\uni}{\mathrm{U}}

\newcommand{\puni}{\mathrm{PU}}

\newcommand{\bn}{\mathbf{n}}

\newcommand{\br}{\mathbf{r}}
\newcommand{\trans}{^{\text{\tiny{$\mathrm T$}}}}
\newcommand{\ii}{\imath}
\newcommand{\ee}{\mathrm{e}}

\newcommand{\se}{_{s(e)}}
\newcommand{\te}{_{t(e)}}

\DeclareMathOperator{\Dir}{\mathscr{D}}

\DeclareMathOperator{\hol}{\mathrm{hol}}
\DeclareMathOperator{\ReppS}{\mathrm{Rep}_{\pS}}

\DeclareMathOperator{\End}{\mathrm{End}}

\DeclareMathOperator{\diag}{\mathrm{diag}}

\DeclareMathOperator{\Adj}{Ad}

\DeclareMathOperator{\Tr}{Tr}

\def\[#1\]{%
  \begin{align}#1\end{align}%
}

\usepackage[margin=1.1949in]{geometry}
\tikzcdset{arrow style=tikz, diagrams={>={Stealth[round,length=4pt,width=4.95pt,inset=2.75pt]}}}

 \title[Loop equations for NCG on quivers]{
The loop equations for\\
 noncommutative geometries on quivers}
 \author[C. I. Perez-Sanchez]{Carlos I. Perez-Sanchez}

  \address{University of Heidelberg, Institute for Theoretical Physics, \newline \indent
  Philosophenweg 19, 69120 Heidelberg, Germany, European Union   \newline \indent
  \hspace{.0cm}\& \newline \indent
  Erwin Schrödinger International Institute for Mathematics and Physics, \newline \indent University of Vienna, Boltzmanngasse 9
 1090 Wien,  Austria, European Union
  }
\email{perez@thphys.uni-heidelberg.de}

\makeatletter

\newcommand*\notocchapter[1]{%
  \if@openright\cleardoublepage\else\clearpage\fi
  \thispagestyle{empty}\global\@topnum\z@
  \@afterindenttrue
  \let\@secnumber\@empty
  \@makeschapterhead{#1}\@afterheading
}

\makeatother

\newcommand{\de}{{e_\circ}}
\newcommand{\deo}{e_\circ}
\newcommand{\bde}{{{\bar{e}_\circ}}}

\begin{document}

\begin{abstract}
We define a path integral over Dirac operators that averages over
noncommutative geometries on a fixed graph, as the title reveals,
using quiver representations.
We prove algebraic relations
that are satisfied by the expectation value of the respective
observables, computed in terms of integrals over unitary groups, with
weights defined by the spectral action. These equations generalise the
Makeenko-Migdal equations---the constraints of lattice gauge
theory---from lattices to arbitrary graphs.  As a perspective, our
loop equations are combined with positivity conditions (on a matrix
parametrised by composition of Wilson loops). On a simple quiver this
combination known as `bootstrap' is fully worked out.  The respective
partition function boils down to an integral known
as Gross-Witten-Wadia model; their solution confirms the solution
bootstrapped by our loop equations.
\end{abstract}
\maketitle%
\section{Introduction}\label{sec:Intro}
Before discussing our problem in its due context, we describe it
aridly, postponing its motivation for Section
\ref{sec:Motivation}. For integers $N$ and $n$ satisfying $N > n > 1$,
fix a polynomial $S\in\C_{\langle 2 n\rangle}=\C{\langle u_1, u_1^*,
  u_2, u_2^*,\ldots, u_n,u_n^*\rangle}$ in noncommutative
$u$-variables satisfying $u_ju_j^*=1=u_j^* u_j$ for $j=1,\ldots, n$.
For a given  polynomial $\beta \in \C_{\langle 2 n \rangle}  $ consider a family of integrals of the type
\[ \label{unitaryintegral} \raisetag{0ex} I_\beta=  \int_{\uni(N)^n}  \Tr \beta(U_1,U^*_1,\ldots, U_n,U_n^*)
\ee^{N \Tr S(U_1,U^*_1,\ldots, U_n,U_n^*)} \dif U_1 \dif U_2 \cdots
\dif U_n,
\]
with each
factor $\dif U_i$ being the Haar measure on $\uni(N)$.  Assuming that
$\Tr S$ is real-valued over the whole integration domain, we derive
the \textit{loop equations}, that is to say, algebraic relations among
the integrals $\{I_\beta\}_{\beta \in \mathcal I } $ parametrised by a
certain family $\mathcal I \subset \C_{\langle 2n \rangle}$.  This
type of integrals has been considered by physicists in the context of
lattice gauge field theory. In mathematics, integrals over the unitary
group are relevant in the context of Weingarten-calculus
\cite{Collins:2003ncs}, developed mainly by Collins and collaborators
(e.g. \cite{Collins:2006jgn,Collins:2020iri}).

\subsection{Motivation: Random matrix theory and noncommutative geometry}\label{sec:Motivation}
Our interest in integrals of the type \eqref{unitaryintegral} emerges
from Connes' noncommutative geometrical \cite{ConnesNCGbook} approach
to fundamental interactions, in which geometric notions are mainly
governed by a self-adjoint operator $D$ named after Dirac. In this
setting, the physical action $S(D)$ is claimed to depend only on (the
spectrum of) $D$ and is known as spectral action
\cite{Chamseddine:1996zu}.  The problem
that motivates this article is the evaluation of the moments that the spectral action yields via
\[
\mathbb E [ h(D) ] = \frac{1}{\mathcal Z}  \int_{\text{Dirac}} h(D) \ee^{-S(D)} \dif D, \label{CM} \qquad \mathbb E [1 ] =1,\quad h(D)\in \R,
\]
for an ensemble of Dirac operators $D$ (the normalisation condition
defines $\mathcal Z$). Of course, this requires to have defined the
measure $\dif D$ on such ensemble, as well as the ensemble itself.  (In the problem originally
formulated in \cite[Sec. 19]{CMbook} the spectral action contains fermions, as it has
been recently addressed in \cite{Khalkhali:2024tyl}, but which we do
not include here.)  \par Part of the relatively vivid interest in the
problem \eqref{CM} during the last decade is due to the reformulation
\cite{Barrett:2015naa} of fuzzy spaces\footnote{We do not aim at a
comprehensive review here, for fuzzy spaces see
e.g. \cite{Steinacker:2007iq} and the works of Rieffel
\cite{Rieffel:2007hv,Rieffel:2007hv,Rieffel:2021ykh} (and references
therein) that address, from diverse mathematical angles, the rigorous
convergence of matrix algebras to the sphere. We are also not
reviewing all the quantisation approaches either; for a
Batalin-Vilkovisky approach: cf \cite{Iseppi:2016olv} for Tate-Koszul
resolutions applied to a model of $2\times 2$-matrices and
\cite{Gaunt:2022elo,Nguyen:2021rsa} for the homological-perturbative
approach to Dirac-operator valued integrals.  } as finite-dimensional
spectral triples. This led to the application of tools related to
random matrix theory
\cite{TRNCG,SAfuzzy,Khalkhali:2020djp,Perez-Sanchez:2020kgq,Perez-Sanchez:2021vpf,Hessam:2022gaw}
that followed to the first numerical results \cite{Barrett:2015foa}.
All these works deal with multimatrix interactions that include a
product of traces (as opposed to the ordinary {potentials in random matrix theory}, namely a
single trace of a noncommutative polynomial). {We elaborate on
the physical interpretation below in Sec. \ref{sec:phys_interpr}}

{Still with a similar aim and yet with other techniques, before our developments,} \rojo{independently}
the Taylor expansion of the spectral action yields
in  \cite[Cor. 19]{vanSuijlekom:PertOpTrace}
a hermitian one-matrix model of the form
\[
V(M) = \sum_{l=1}^\infty \sum_{i_1,i_2,\ldots,i_l} F_{i_1,i_2,\ldots, i_l} M_{i_1,i_2}
M_{i_2,i_3}\cdots M_{i_{l},i_1},  \text{ with }F_{i_1,i_2,\ldots, i_l}  \in \R\,. \] 
This series was shown in  \cite{vanNuland:2021otn} to be convergent under certain conditions and, combining some elements of  \cite{Connes:2006qj} with own techniques, to possess a neat reorganisation in terms of a series expansion in
universal Chern-Simons forms
and Yang-Mills forms integrated against $(B,b)$-cocycles that do depend on the geometry.
Each monomial of the model $V(M)$ above breaks unitariness and thus
goes beyond the solved generalisations \cite{zbMATH07650768}
of the Kontsevich matrix model \cite{KontsevichModel} (in which unitary invariance
is broken only by the propagator) known as Grosse-Wulkenhaar model \cite{GW12}.
\\

These two independent approaches portend a symbiosis between
random matrix theory and noncommutative geometry.  Both the multiple
trace interactions and the unitary-broken interactions could motivate
(if they have not yet) new developments in random matrix theory.  And
vice versa, the path-integral quantisation \eqref{CM} of
noncommutative geometries seems hopeless without the intervention of
random matrix theory\footnote{The only alternative known to the author
is the use of Choi-Effros operators systems
\cite{Connes:2020ifm,Tolerance} (cf. also \cite{Tolerance_Lizzi}) that
emerge when one assumes (or rather, when one accepts) that only a
finite part of the Dirac spectrum is measurable. The price to pay is
nonassociativity.  }.  \par

\subsection{Ensembles of unitary matrices in noncommutative geometry}
The interaction between the two aforementioned disciplines has taken place in spaces of hermitian matrices.
One of the novelties of this article is that the
integrals over Dirac operators defined here  boil down  to
ensembles of unitary matrices instead (they are also unitary-invariant,
like ordinary hermitian matrix ensembles, but
unitary ensembles integrate over unitary random matrices). These
can be considered as an approach to average over
 `noncommutative geometries
on a graph'.
When the graph is provided with
additional structure,
it might be grasped as a discretisation of space.
For instance, edges would carry a representation while vertices
equivariant maps; at least so in the
  spin network approach. Here, we refrain from including information
  associated to  gravitational degrees of freedom and
  address exclusively the  problem of gauge interactions.
  The background geometry is therefore fixed
  and the finiteness of the unitary groups appearing is not
  a shortage of the theory; as a caveat,
  they are not to be interpreted as a truncation of infinite-dimensional
  symmetries (but to be compared with the unitary structure group of
  Yang-Mills, for example). \\

Representation theory does still play a role, but rather
in the context of quiver representations in  a certain category
that emerges from noncommutative geometry, as
exposed in \cite{NCGquivers} after the pioneering\footnote{\cite{NCGquivers} shows that, unfortunately, \cite{MvS} errs twice: first,
the characterisation of the gauge group (that parametrises equivalences of representations)
given by Marcolli-van Suijlekom's theory yields a group that is too big, cf. \cite[Rem. 3.18]{NCGquivers}; secondly,
the morphism structure that they imposed on the spectral triples category
dooms the Higgs field constructed by them to be a constant, as shown in \cite[Sec. 5.3]{NCGquivers}.
\cite{NCGquivers} provides a dynamical Higgs and the correct quiver gauge group along with new results.}
ideas of \cite{MvS}.
\\

We can now restate the aim of this article as follows:
\begin{quotation}
\textit{Define a partition function for noncommutative geometries on a
  graph---that is, define a measure over all `compatible' Dirac
  operators---and prove algebraic relations that the respective
  observables shall satisfy.  Such quantities have the form $I_\beta$
  as in eq. \eqref{unitaryintegral} and are called Wilson loops
  (although not each $I_\beta$ is a Wilson loop on a given graph).}
\end{quotation}
Proper definitions follow in the main text.  Such relations generalise
the Makeenko-Migdal equations, i.e. the loop equations in lattice gauge
theory.  After introducing the setting in Section
\ref{sec:QuiversNCG}, we prove the main result in Section
\ref{sec:main} and conclude with a fully worked-out application that
mixes the loop equations with positivity conditions of a certain
matrix (`bootstrap') in Section \ref{sec:Applications}.

\subsection{Physical motivation}\label{sec:phys_interpr}
From the physics perspective\footnote{The author thanks a referee for
comments that lead to writing this paragraph.\label{fn:Ref2}}, most of
the references
\cite{ Barrett:2015foa, TRNCG,SAfuzzy,Khalkhali:2020djp,Perez-Sanchez:2020kgq,Perez-Sanchez:2021vpf,Hessam:2022gaw}   that address matrix or fuzzy geometries from the viewpoint of random Dirac operators,
focus on---or, at least, were motivated by---quantum gravity, while gauge interactions are absent or
still classical as in
\cite{Perez-Sanchez:2021vpf}. Diametrically opposite
to the previous references, in the present article,  we keep  an oriented graph (the quiver) fixed.
If an interpretation is really wished, such quiver will
serve as a `background' and one could grasp it as a deterministic surrogate for a discrete spacetime.
However, since the
quiver still does not include information regarding a
metric or a connection,
gravity should be understood as being `turned off'. Instead
of decorating the quiver
(e.g. by enriching its
 edges with coloring, group-labels or representation data like tensor models \cite{GurauRyan},
 group field theory  \cite{Oriti} or  spin networks \cite{Baez} respectively do in order
to eventually model gravitational degrees of freedom)
our main aim in this article is to formulate, as a partition function that integrates over Dirac operators, quantum gauge theories with the immediate possibility to incorporate a dynamic Higgs scalar on a deterministic
background discretised by a quiver. (That a Dirac operator
is   able to describe gauge, Higgs and gravity degrees of freedom is known from classical noncommutative geometry results, e.g.
 \cite[Ch. 9-11]{CMbook}. As a caveat, although these multiple facets
 of the Dirac operator allow us to focus on the one or on the other sector,
there does not seem to exist a way to recover [e.g. by choosing a specific quiver] fuzzy geometries
 when we will `turn gravity on'. We are formulating
 another mathematical model that is explicitly independent from
 fuzzy geometries, but which shares the aim of studying random Dirac operators,
even if these are used for different physical theories.)

\section{Quiver representations and noncommutative geometry} \label{sec:QuiversNCG}
We call \textit{quiver} $Q$ a directed multigraph.  Since $Q$ is
directed, there are maps $s,t: Q_1 \rightrightarrows Q_0$ (from the
edge-set $Q_1$ to the vertex-set $Q_0$) determining the vertex $s(e)$
at which an edge $e$ begins, and the one $t(e)$ where it ends.
\textit{Multiple edges} $e,e' \in Q_1$ and \textit{self-loops} $
o_v\in Q_0$ at a certain vertex $v\in Q_0$ are allowed, namely $ \{
s(e) , t(e) \} = \{s(e'),t(e')\}$ as sets, and $s(o_v)=t(o_v) =v $,
respectively.  \\

With a quiver $Q$ one can associate the \textit{path category} whose objects are $Q_0$; the
morphisms $\hom_{PQ}(v, w)$ are the \textit{paths} $\gamma$ from $v$ to $w$,
namely edge-sequences $\gamma=(e_1 e_2\cdots e_n)$ with $e_1,\ldots,
e_n\in Q_1$ and $s(\gamma)=s(e_1)=v$, and $t(\gamma)=t(e_n)=w$ as well
as $t(e_j)=s(e_{j+1})$ for $j=1,\ldots, n-1$. The set of paths
on $Q$ is denoted by $P Q$ and slightly abusing on notation, so will be
also denoted the free or path category
associated to $Q$. \\

We shall write
$\gamma:v \to w$ if $v=s(\gamma)$ and $w=t(\gamma)$ and call
$\ell(\gamma)=n$ the \textit{length} of $\gamma$.  The path $\gamma$
with reversed order is denoted by $\bar \gamma=(e_n e_{n-1}\cdots
e_2e_1)$ (not to be confused with the inverse morphism of $\gamma$).
Obviously, unless otherwise stated, paths are directed, but it will
prove useful to consider also paths in $\Gamma Q$, the underlying
graph of the quiver ($Q$ with forgotten orientations). If
$s(\gamma)=t(\gamma)$ we say that a path $\gamma$ is a
\textit{loop}. The space of loops\footnote{We comment for sake of
completeness, that the space of endomorphisms
$\Omega_v(Q)=\hom_{PQ}(v,v)$ has as identity the constant zero-length
path, which does play a role in the theory of path algebras while
constructing an equivalence between the category of representation and
modules of the path algebra \cite{bookQuivRep}, but which we do not
need here explicitly.} at $v$, is denoted here $\Omega_v(Q)$, that is
$\Omega_v(Q)=\hom_{PQ}(v,v)$, and $\Omega Q$ will denote the space
$\cup_{v \in {Q_0}} \Omega_v(Q)$ of all loops.  \\

One uses the terminology `quiver' for a multidigraph $Q$ when one wants to represent $Q$: To wit,
given a category $\mathcal C$, a $\mathcal C$-\textit{representation} of $Q$ is
by definition a functor from $PQ$ to $\mathcal C$.%

\subsection{The spectral triple associated to a quiver representation}

We restrict the discussion to finite dimensions and introduce the
setting of \cite{NCGquivers}. We dedicated Section \ref{sec:illustration}
to examples of the new constructions that appear here.
By definition, an object in the
category $\pS$ of \textit{prespectral triples} is a pair $(A,H)$ of a
unital $*$-algebra $A$ faithfully $*$-represented, $\lambda: A
\curvearrowright H$, in an inner product $\C$-vector space $H$
($*$-represented means here, that $\lambda(a^*)$ is the adjoint
operator of $\lambda(a)$ for all $a\in A$).  A morphisms in $
\hom_{\pS} ( A_s,H_s \, ;\, A_t,H_t ) $ is a couple $(\phi,U)$ of an
involutive unital algebra map $\phi: A_s \to A_t$ as well as a unitary
map $U: H_s \to H_t$. As part of the definition, a morphism should in
addition satisfy $ U \lambda_s (a) U^* = \lambda_t [ \phi (a) ]$ for
all $a\in A$. \\[.12ex]

\noindent
\begin{minipage}{.74\textwidth}
In other words, a $\pS$-representation of $Q$ associates with each
vertex $v $ of $Q$ a prespectral triple $(A_v,H_v) \in \pS$ and with
any path $\gamma: v\to w$ a morphism $(\phi_\gamma,\hol \gamma ) :
(A_v,H_v) \to (A_w,H_w) $ in such a way that if $\gamma = (e_1\cdots
e_n)$, then $\hol \gamma= U_{e_n} \cdots U_{e_1}$ and $\phi_\gamma=
\phi_{e_n}\circ \phi_{e_{n-1}} \circ \cdots \circ \phi_{e_1}$, where
$\phi_{e_j } : A_{s(e_j)}\to A_{t(e_j)}$ and $U_{e_j}: H_{s(e_j)} \to
H_{t(e_j)}$ form a $\pS$-morphism.  We refer to $\hol \gamma$ as the
\textit{holonomy} of $\gamma$.  (If $\gamma$ is not a loop,
\textit{parallel transport} would be the precise term; for sake of
notation, we call this `holonomy' too.)  \\[.5ex]
\end{minipage}~
\begin{minipage}{.23\textwidth} \vspace{-1ex}
\[\qquad
 \notag \includegraphics[width=4cm]{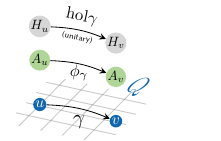} \]
\end{minipage}

\par If two vertices are connected by a path $\gamma$, notice that $
\hol \gamma$ is a unitary matrix and $\dim H_{s(\gamma)} =\dim
H_{t(\gamma)}$.  If $Q$ is connected, there might be no (directed) path
between two given vertices $v$ and $w$; it is however easy---if necessary after inverting some subpaths of a path $\tilde \gamma$ in
$\Gamma Q$ that connects $v$ with $w$---to establish the constancy of
the map $Q_0\ni v\mapsto \dim H_v:=N$; we call such constant $N=\dim
R$, \textit{the dimension of the representation $R$}, somehow
abusively. \\

A \textit{spectral triple} $(A,H,D)$ is a prespectral
triple $(A,H)$ together with a self-adjoint element $D \in \End(H)$,
referred to as \textit{Dirac operator}.  (This terminology comes from
the non-trivial statement that $D$
is the spin geometry Dirac operator \cite{Connes:2008vs},
if certain operators are added to the
[in that case, infinite-dimensional] spectral triple and if, together
with $D$, such operators satisfy a meticulous list of axioms; see also
\cite{WvSbook} for an introduction geared to physicists).

\begin{remark}\label{rem:differences}
As a side note, it is possible to compute the space of all
$\pS$-representations of $Q$. It was proven in \cite{NCGquivers} that
such space---which in fact forms the category of representations---can
be described in terms of products of unitary groups subordinated to
combinatorial devices called Bratteli networks
(Sec. \ref{sec:dDandZ}).  At this point, it is important to observe
that, in stark contrast with ordinary $\textsc{Vect}_\C$-quiver
representations, providing labels to the vertices is not enough to
determine a $\pS$-quiver representation.  The lifts of whole paths
should exist, and this requires the compatibility of the maps $\phi_v$
at all vertices $v$, which in turn is what the so-called Bratteli
networks guarantee (concretely unital $*$-algebra maps for $M_m(\C)
\to M_{n}(\C)$ for $m>n$ do not exist, and if a representation yields
$A_{s(e)} =\M m $ and $A_{t(e)} =\M n$ for some edge $e$, a lift
fails, cf. \cite[Ex. 3.19]{NCGquivers}).  Despite this, we denote
representations of quivers as $R=\{ (A_v,H_v), (\phi_e,U_e) \} _{v\in
  Q_0,e\in Q_1}$ instead of $R=\{ (A_v,H_v), (\phi_\gamma,\hol \gamma)
\} _{v\in Q_0,\gamma \in \Omega Q}$, meanwhile under the tacit assumption that
lifts of whole paths exist. A characterization follows in next the section.
\end{remark}

We associate now a spectral triple to a given $\pS$-representation $R$ of a connected quiver $Q$,
$R=\{ (A_v,H_v), (\phi_e,U_e) \} _{v\in Q_0,e\in Q_1}$.  We define the Dirac operator associated to $R$ as the
matrix $D_Q(R) \in M_{ \# Q_0} (\C) \otimes M_N(\C) $ with matrix
entries $[D_Q(R)]_{v,w} \in M_N(\C)$ in the second factor given by
 \[ \label{DiracQ}
[D_Q(R)]_{v,w} = \bigg(
\sum_{e \in s\inv (v) \cap t\inv (w)}
U_e\bigg) +\bigg(
\sum_{e \in t\inv (v) \cap s\inv (w)}
U_{e}^* \bigg)  \qquad (v,w\in Q_0).
\]
This Dirac operator is not a guess. It is the correct one, since it yields lattice gauge theory
if $Q$ is a lattice, cf. \cite[Thm. 5.2]{NCGquivers}.
By construction, this operator is self-adjoint, and crucially for our
purposes, the objects form a spectral triple,
\[ \label{spectraltripleQ}
\big (A_Q(R), H_Q(R), D_Q(R)  \big ) =
\bigg ( \bigoplus _ {v\in Q_0} A_v, \bigoplus _ {v\in Q_0} H_v, D_Q(R) \bigg).
\]
This direct sum is dictated by the equivalence of categories
of path algebra modules with that of quiver representations (for
the target category $\pS$, see \cite[Prop. 3.10]{NCGquivers}).
In Definition \ref{def:space_of_DiracsB}, the space of Dirac operators
will be (below, carefully) defined as a Haar measure for all the $U_e$ variables above
in eq. \eqref{DiracQ}. The correct definition needs to fix certain data
that are introduced in the next section.


\subsection{The spectral action}
\label{sec:SA}

Given a polynomial $f(x) = f_0 + f_1 x^1 +f_2 x^2 +\ldots+ f_d x^d $
in real variables $f_0,f_1,\ldots, f_d \in \mathbb R$, and a quiver
representation, the spectral action on a quiver reads $S( D) = \Tr_H
f(D) $, where we abbreviate $D=D_Q(R)$ and $H=H_Q(R)$.   It is possible to compute the
spectral action as a loop expansion in terms of \textit{generalised
  plaquettes} $\gamma$ as follows
\[ \label{loopexpSA}
\Tr_H f(D)=   \sum_{k=1}^d f_k \,\sum_{v \in Q_0} \,\,
\sumsub{ \gamma \in \Omega_v(Q) \\  \ell(\gamma)  =k } \Tr \hol \gamma,
\]
where $\Tr$ in the rhs is the trace of $M_N(\C)$ with $\Tr 1=\dim
R=N$.  The proof of eq. \eqref{loopexpSA} is given in
\cite{NCGquivers}, but the reader will recognise this formula as a
noncommutative generalisation of the following well-known fact in
graph theory: if $C_G$ denotes the adjacency matrix of a graph $G$,
then the number of length-$n$ paths in $G$ between two of its
vertices, $i$ and $j$, is the entry $[C_G^n] _{i,j}$ of the matrix
$(C_G)^n$.\par

\subsection{The measure on the space of Dirac operators and the partition function}
\label{sec:dDandZ}
Now we break down the space\footnote{
Several new concepts appear in this subsection.
For the readers who prefer an exposition that mixes examples with and definitions,
we encourage a non-linear reading, alternating
Sec. \ref{sec:dDandZ} with the examples of Sec. \ref{sec:illustration} (or even the latter first).} of $\pS$-representations of $Q$,
\[\ReppS(Q) := \{ \text{functors } PQ\to\pS \} . \label{beliebig}\]

Let $A_v=\oplus_{j=1}^{l_v} \M {n_{v, j}} $ denote the algebra
associated by $R$ to the vertex $v$ (so $l_v$ is the number of simple
subalgebras of $A_v$).  Let $r_{v,j}$ be the multiplicity of the
action of the factor $\M{n_{v,j}} \subset A_v$ on the Hilbert space
$H_v$, that is $H_v= \oplus_{j=1}^{l_v} \C^{ r_{v,j}} \otimes
\C^{n_{v,j} }$ where $\M{n_{v,j}} $ only acts non-trivially on
$\C^{n_{v,j} }$ via the fundamental representation. These integers are
not arbitrary, since clearly the totality of the $\{ n_{v,1},\ldots,
n_{v,l_v} \}_v$ should be such that unital $*$-algebra maps between
vertices connected by an edge exist.  The next definition, reformulated from \cite{NCGquivers},
captures this requirements. \par

\begin{definition}\label{def:BratteliNetz}
A \textit{Bratteli network} $B$ on a connected quiver $Q$
consists of the following data:
\begin{enumerate}
\vspace{.51ex}\item an integer $l_v > 0$ for each vertex $v\in Q_0$

 \vspace{.51ex}\item a $  l_v$-tuple  $\mathbf r_v \in \N^{l_v}$ for each vertex

\vspace{.51ex}\item another $l_v$-tuple $\mathbf n_v\in \N^{l_v} $for each $v\in Q_0$

 \vspace{.51ex}\item for each edge $e \in Q_1$,
 a matrix $C_e \in  M_{l_{s(e)} \times l_{t(e)}} (\mathbb Z_{\geq 0})$
 such that
 \[ \label{transmatrix}
 \br_{s(e) } = C_e \br_{t(e) }
 \quad \text{ and }\quad
\bn_{t(e) } = C\trans_e \bn_{s(e) }.
 \]
\end{enumerate}
\end{definition}
For sake of notation, we denote Bratteli networks with the variables
$B$ or $ (\bn,\br) $ leaving the rest of data implicit. If $\langle
\mathbf a , \mathbf b \rangle= \sum_i a_i b_i$ is the standard
bilinear form on $\Z^\infty \times\Z^\infty$, it is essential to
observe that Conditions \eqref{transmatrix} guarantee that
\[
\langle \bn\se,\br\se \rangle
=
\langle
\bn\se, C\trans _e \br\te
\rangle
=\langle
C_e \bn\se,  \br\te
\rangle
=\langle
\bn\te,  \br\te
\rangle 
\label{constantN}, \qquad e\in Q_1,
\]
is a constant integer $N$, whenever a quiver is connected. A quiver
$\pS$-representation determines a Bratteli network in the following way.
Given any $R=\{ (A_v,H_v),$ $ (\phi_e,U_e) $$\} _{v\in
  Q_0,e\in Q_1}$ $  \in \ReppS(Q)$,
  by the discussion below eq. \eqref{beliebig},
  there exist integers $l_v$ and $ n_{v, j}, r_{v, j}$ for $j=1,\ldots, l_v$
  such that
  $A_v=\oplus_{j=1}^{l_v} \M {n_{v, j}}$
and $H_v =  \oplus_{j=1}^{l_v} \C^{ r_{v,j}} \otimes
\C^{n_{v,j} }$. The associated Bratteli network is composed by
$(\bn_v,\br_n)$ with $ \bn_v= (n_{v,1},\ldots, n_{v,l_v})$,
$ \br_v= (r_{v,1},\ldots, r_{v,l_v})$  (notations coincide on purpose, and $N$ is the
dimension of the representation). They are easily seen to verify eq. \eqref{transmatrix}
due to the existence of the $*$-algebra morphisms $\phi_e$, $e\in Q_1$.\\

\noindent
The next question arises:
\begin{quote}
 \textit{what is missing a Bratteli network $B$ in order to determine a quiver $\pS$-representation?}
\end{quote}
In the light of the spectral triple $ ( \oplus _ {v\in Q_0} A_v,
\oplus _ {v\in Q_0} H_v, D_Q(R) )$ that is associated to a quiver
representation, since a Bratteli network is equivalent to the first
pair of objects, $(\oplus _ {v\in Q_0} A_v, \oplus _ {v\in Q_0} H_v)$,
the relevant answer is that the missing piece is the Dirac operator
associated to the quiver.  They exist in abundance and we are
interested in their probability distribution.

\par

\begin{definition}\label{def:space_of_DiracsB} Given a Bratteli network
$ B=\{ A_v,H_v \} _{v\in Q_0}$ on a connected quiver $Q$, the
  \textit{space of Dirac operators} $\Dir ( { B}) $ is defined as the
  set of $\pS$-maps between vertices
\[    D=\big \{ (\phi_e,U_e) \in \hom_\pS(
A\se ,H\se \,;\,\, A\te,  H\te
  ) \big \} _{e\in Q_1} \label{C}  \]
that complete $  B$ and make it a $\pS$-representation of $Q$, that is
\[
\Dir ( {  B}) : =
\{\,    D \text{ as in } \eqref{C}
   \mid   \, (    B ,   D ) \in \ReppS (Q)  \}. \notag
\]
\end{definition}

\begin{figure}[htb]
\includegraphics[width=7cm]{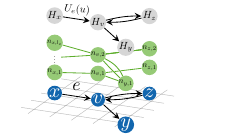}
\caption{Illustration of a quiver on the plane (when color is available, stressed with blue)
and some $\pS$-representation data. Here, a green node tagged with $n$ represents $\M n$;
vertically aligned  nodes should be understood as  their direct sum. Green edges represent the embedding of the
matrix algebras associated to the source-vertex into the matrix
algebra associated to the target-vertex of any edge $e$ (the consistency
conditions like $n_{y,1}=  n_{v,1}+2 n_{v,2}$ implied by such lines are implicit). The Hilbert spaces at the vertices
are depicted on the third layer (above) with gray. In particular, $U_e:H_x\to H_v$ depends
on the lower-case $u$-variables $ u \in \prod_{j=1,\ldots, l_{v}} \uni(n_{v,j})$ above $t(e)=v$ as mentioned in Remark \ref{rem:Us}. This means that, after the Bratteli network is fixed (corresponding
to specifying the labels on the green and gray nodes) the only information left to be provided
in order to fully specify a $\pS$-representation is an element of
$ \prod_{e \in Q_1} \prod_{j=1,\ldots, l_{t(e)}} \uni(n_{t(e),j})$. This space parametrises all
possible Dirac operators.
\label{fig:Us}}
\end{figure}

Once labels to the vertices are consistently assigned by the Bratteli network $B$, the possible
labels of an edge $e$ are parametrised\footnote{The reader will note
that we do not include the minimal amount of information in each group
at the edges.  The origin of the projective groups $\puni(n)$ is that
$\uni(n)$ acts via the adjoint action.  } by $\prod_{j=1,\ldots,
  l_{t(e)}}\uni(n_{t(e),j})$ \cite[Lemma 3.12]{NCGquivers}.  Then
\[ \label{DirSpace}
\Dir ( {  B})
= \coprod_{e\in Q_1}
\prod_{j=1,\ldots, l_{t(e)}}\uni(n_{t(e),j}) .
\]
\begin{remark} \label{rem:Us}
Without aiming at repeating the proof given there, let us explain
here why these unitary groups $\uni(n_{t(e),j})$ appear.
If $\phi: \M m \to \M n$ is a $*$-algebra morphism,
then $m$ divides $n$ ($r\cdot m=n$)
and is given by $\phi (a)=a\oplus \cdots \oplus a$ (with $r$ summands) up to a unitary matrix $u$,
i.e. the most general morphism is then
 $\phi_u(a)=  u a u^*  \oplus \cdots \oplus u a u^*$ for a certain $u\in \uni(n)$.
When this situation is analysed in full generality for direct sums of simple matrix algebras,
as shown in
 \cite[Lemma 3.12]{NCGquivers} inspired in \cite{MvS} (who
 were inspired by Bratteli),
if $\phi: \oplus_i ^{l_s} \M {m_i} \to \oplus_j ^ {l_t} \M {n_j}$,
 is a $*$-algebra morphism, then so is $\Adj u\circ  \phi $,
 where $u=(u_1,\ldots, u_{l_t}) \in \prod_{j=1,\ldots, l_t} \uni(n_{j})  $.
The more complex situation in \eqref{DirSpace} is due to the dependence of these
unitary groups on the vertices $t(e)$ where all the edges $e\in Q_1$ end, see
Figure \ref{fig:Us}.
\end{remark}

The overlapping notation was then on purpose, as $D \in \Dir ( { B})$,
and $D_Q(R)$ as the Dirac operator of the spectral triple associated
to a quiver $\pS$-representation, entail the same information. This in
turn motivates the following measure.

\begin{definition}\label{def:measure}
Given a Bratteli network $ B$ on a connected quiver $Q$, we define the
\textit{Dirac operator measure} $\dif D$ on the space of Dirac
operators $\Dir( B)$ by
\[
\dif  D  := \prod_{(v,w) \in Q_0\times Q_0 } \dif [D_Q(R)]_{v,w}, \quad
\text{ where } \quad\dif [ D_Q(R)]_{v,w}  := \prod _{ e \in s\inv (v) \cap t\inv (w) }
\prod_{j=1}^{l_{t(e)}}
\dif u_{e,j},
\]%
\noindent
being $\dif u_{e,j}$ the Haar measure on $ \uni ( { {n_{t(e),j} }} )$,
where $u_{e,j}$ sits in the matrix $U_e$ associated to $e$ by $R$ in
the respective block-diagonal entry in
\[ \label{blockembedding}
 U_e =U_e\big(\{ u_{e,i}\}_i\big)=\diag (
        1_{r_{t(e),1 } } \otimes u_{e,1}
,      1_{r_{t(e),2}} \otimes u_{e,2} ,\ldots,
      1_{r_{ t(e) ,l_{t(e)} }} \otimes u_{e, l_{t(e)}} ).\]
\end{definition}%

\begin{definition}\label{def:partfunc}
Given a Bratteli network $B$ on a quiver  $Q$,
the \textit{partition function}
reads
\[ \label{partfunc}
\mathcal Z_{Q,  B  }(f)= \int _{\Dir(  B)} \ee^{-N  \Tr_H f(D) } \dif D, \qquad
\]
 where $D \in \Dir (B)$
 complements the initial Bratteli network
 making of it a representation $R=(B,D)$
 of dimension $ \dim R=N$ given by the integer
 \eqref{constantN}. In the Boltzmann weight, the
 spectral action $\Tr_H f(D)$ is given by eq. \eqref{loopexpSA}.
\end{definition}

\begin{remark} Some remarks related to the meaning of the partition function:

\begin{enumerate}
\vspace{.51ex}\item The Dirac operator measure $\dif D$ is the Haar measure on $\prod_{e \in Q_1 }
\prod_{i=1}^{l_{t(e)}} \uni ( {{n_{t(e),i} }})$
$\hookrightarrow\uni(N)^{\# Q_1 }$ since
$ \langle \bn_v,\br_v  \rangle = \sum_{i=1,\ldots, l_v} r_{v,j} \times n_{{v,i}} = N $
 holds at each vertex, by
eq. \eqref{constantN}.
\vspace{.51ex}\item
 In the gauge theory picture,  $Q$ is a coarse set of data for the base manifold (of a principal bundle).
 A Bratteli network on $Q$ predetermines a `local field of gauge groups', that is $Q_0 \ni v\mapsto \mathcal
U(A_v)$. The holonomies of paths will therefore gather unitary matrices that
can be multiplied thanks to the embedding
\eqref{blockembedding}. It would be interesting
to explore whether the present structures
relate to lifts of Krajewski diagrams (that classify
finite spectral triples \cite{Krajewski:classif,PaschkeSitarz}) in the sense of
\cite{Masson:2022ppv} in some
special cases of one or both theories.

\vspace{.51ex}\item Due to (2) of this remark and because of the
previous identification of a Bratteli network $B$ with fixed data  $(A_Q(R),H_Q(R), \,\balita\,)$
of the spectral triple in \eqref{spectraltripleQ},
if $S(D)$ the spectral action  \eqref{loopexpSA},
the partition function in  \eqref{partfunc} is of the form
\[
\mathcal Z_{A_Q,H_Q}(f) = \int_
{\substack{\phantom{nix} \\D \text{ makes } (A_Q,H_Q, D) \\ \text{into a spectral triple} }}
 \ee^{-N S ( D )} \dif D.
\]
\vspace{.1ex}\item According to the definition of $D_Q(R)$ in
eq. \eqref{DiracQ}, the Dirac operators' entries determine self-adjoint matrices
$\mathsf A_{e} \in (A\te )_{\sa} $, interpreted as connections, given by  $\mathsf A_{e} =
\diag(\mathsf a_{e,1},\mathsf a_{e,2},\ldots, \mathsf a_{e,l\te}) $ along the edges
by $u_{e,i}=:
\exp(\sqrt{-1} \mathsf a_{e,i})$ for $i=1,\ldots, l\te$, cf. eqs \eqref{blockembedding}.
 \vspace{.51ex}\item For fixed $N$, the partition function $\mathcal Z _Q= \sum_{R \text{
    $\pS$-rep of $Q$} }^{\dim R=N} \mathcal Z_{Q,R}$ is also an
interesting quantity, or even more so the sum over a class of quivers
$Q$ encoding different background geometries, $\mathcal Z=\sum_Q
\sum_{R \text{ $\pS$-rep of $Q$} }^{\dim R=N} \mathcal Z_{Q,R}$. For
the moment we content ourselves with the partition function
\eqref{partfunc} for a fixed Bratteli network $B$ and a fixed quiver
$Q$.
\end{enumerate}

\end{remark}

 \begin{definition}
\label{def:Wilson}
  For any $ \beta \in \Omega (\Gamma Q)$, a \textit{Wilson loop}\footnote{We
  refer both to $\hol \beta$ and to $\mathbb E [ \Tr \hol \beta]$
  ambiguously as Wilson loops.} is by definition
\[
\mathbb E [ \Tr ( \hol \beta) ]:=
\frac1{\mathcal Z_{Q,  B}} \int_{\Dir(  B)}
\Tr \hol (\beta ) \ee^{- N S(D)}
\dif D.\notag
\]
\end{definition}

If in the large-$N$ limit factorisation of traces (freeness) holds, then all
observables $ \mathbb E [ \prod_i^{\text{finite}}\Tr ( \hol \beta_i) ]$, $\beta_i \in \Omega (\Gamma Q)$, for the path integral  $\mathcal Z_{Q,  B}$ can be computed
computed from the fundamental (in the sense of single-trace) observables
$\mathbb E [ \Tr ( \hol \beta) ]$. A more common notation for this integral is $
 \langle  \Tr ( \hol \beta) \rangle$.
\vspace{-3ex}
\subsection{Illustrating the previous section} \label{sec:illustration}
Let us pick an example quiver $Q= {\includegraphics[width=8ex]{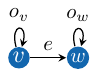}}
$, whose self-loops are denoted by $o$-variables, as before. The concepts
introduced the last subsection are exemplified in the following list.
\begin{enumerate}
 \item \textit{Bratteli network, Def. \ref{def:BratteliNetz}}.
 On $Q$ as above, an example of data of a Bratteli network is
 \[ \notag
 l_v & = 2   &&& l_w &=1    &&& C_e &= (2,1) \trans  \\
 \bn_v & = (3,2)\trans &&& \bn_w&= 8 &&& C_{o_v} &=\diag( 1,1)  \notag \\\notag
 \br_v & = (4,2)\trans &&& \br_w&= 2 &&& C_{o_w} &= 1.
  \]
\item \textit{Why is $B$ a Bratteli `network'?} In the illustration
   an integer $n$ in a green (or circular)
  nodes over a vertex
   represents the simple algebra $\M n$.
 The whole algebra
   associated to the vertex  is the sum over all green circles above it.
   Inside gray rectangles the Hilbert spaces acted on by each simple subalgebra
   are represented;  the non-trivial action takes place only on the second factor.
  \[ & \raisebox{-15ex}{\includegraphics[width=.4\textwidth]{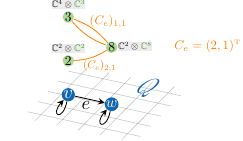}} &&&
H_v&=(\C^4\otimes \C^3) \oplus( \C^2\otimes \C^2 )  \notag \\[-14.2ex]
  & &&& A_v& =\M 3 \oplus \M 2  \notag \\
      &  &&&  H_w&=\C^2\otimes \C^8 \nonumber \\
       &&&& A_w &= \M 8  \notag  \\ &&&&& \notag   \]
 The network arises when all the lines  that $\{C_e\}_{e\in Q_1}$ represent are composed.
 The $C$-matrices associated to the self-loops are the identity and therefore not worth depicting.
  Each unital $*$-algebra map is given
  by block embeddings of the simple algebras into the target algebra  (up to unitary conjugations that parametrise the
  space of Dirac $\Dir(B)$ operators for $B$ below). The entry $(C_e)_{i,j} \in \Znn$ represents how many blocks from the $i$-th factor of $A\se$
  are embedded into the $j$-th factor of $A\te$. For this example, $\phi_e: A_v \to A_w$ is
  $ \phi_e(a,a')=\diag(a,a,a')$, $a\in\M 3$, $a' \in \M 2$. This way the  network emerges,
  which is named after Bratteli due to his work on AF-algebras \cite{BratteliDiags}.
  The information associated to each edge is known as \textit{Bratteli diagram},
  but a Bratteli network is not an arbitrary labelling of edges by Bratteli diagrams.
  They should be also composable and this is guaranteed by the conditions
  that Def. \ref{def:BratteliNetz} imposes on the labels of the vertices.
 \vspace{.51ex}
  \item \textit{Space of Dirac operators, Def. \ref{def:space_of_DiracsB}.}
If $B$ is the previous data,
the space of Dirac operators corresponding to $B$ is $
\Dir(B)= [
\uni(2)\times \uni(3) ]_{o_v} \times \uni (8)_e \times \uni(8)_{o_w}
$
where the subindices refer to the edge that the groups label.

 \vspace{.51ex}\item \textit{How a Bratteli network and a Dirac operator determine a
 quiver $\pS$-representation and the spectral triple for the quiver}.
 The representation $R$ of $Q$ corresponding to $B$ and to an
 element in $(u,u',u'',u''')\in \Dir(B)$
is determined by the following labels of vertices and edges:
 \[
 U_{o_v} & =  \bigg( \begin{matrix}
                    1_4 \otimes u & 0  \\ 0 & 1_2 \otimes u'
                    \end{matrix}
   \bigg) & U_e &= 1_2 \otimes  u''&
 U_{o_w} & =  1_2 \otimes  u'''\,\,. \label{exampleembedings}
 \]
 These, in turn, determine the spectral triple of the eq. \eqref{spectraltripleQ}, namely
\[\notag
\big [ A_Q,H_Q,D_Q(R) \big] =
\bigg[A_v\oplus A_w, H_v\oplus H_w,
\bigg( \begin{matrix}
  \varphi_v   &  U_e  \\
  U^*_e  &  \varphi_w
            \end{matrix} \bigg) \bigg]\,,
\]
whose Dirac operator is constructed according to eq.
\eqref{DiracQ}. The entries abbreviated
 $\varphi_{v} = U_{o_v} + U_{o_v}^* $
and  $\varphi_{w} = U_{o_w} + U_{o_w}^* $ are (hermitian) matrices,
and the four entries are square matrices of size $\dim R=16$, as they should be.

\vspace{.51ex}\item  \textit{Spectral action, Eq. \ref{loopexpSA}.} Choosing $f(z)=z^4$,
the spectral action reads
\[ \Tr_H  f(D) =          \Tr_H \bigg[ \begin{matrix}
  \varphi_v   &  U_e  \\
  U^*_e  &  \varphi_w
            \end{matrix} \bigg]^4
               = \Tr q(\varphi_v) + \Tr q(\varphi_w)   +  4  \Tr ( \varphi_v U_e \varphi_w U^*_e ) .
      \qquad \notag
      \]
      in terms of $q(z):=z^4+4z^2 + 1$. One arrives at this expression by counting paths on $\Gamma Q$.

\vspace{.51ex}
\item  The      \textit{Dirac operator measure, Def. \ref{def:measure},} is the Haar measure on $\Dir(B)=\uni(2)\times\uni(3)\times \uni(8)^2$.

\vspace{.5ex}

\item       \textit{Partition function, Def. \ref{def:partfunc}.} Taking into account the embeddings
\eqref{exampleembedings},
\[ \notag
\mathcal Z_{\!\runterhalb{\includegraphics[width=6ex]{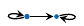}}\!\!,\,B}
&=
\int _{\Dir(B)}
\ee^{ - \dim R \Tr_{H} (D) }
\dif D \\
& =\notag
\int _{\uni(2)\times\uni(3)\times \uni(8)^2}
\ee^{-16 \cdot \Tr [ q(\varphi_v)  +q( \varphi_w )+ 4 \varphi_v U_e \varphi_w U^*_e ] }
 \dif u\,  \, \dif u'\,  \dif u''\,  \dif u'''\,.
\]
 \vspace{.51ex}\item \textit{Wilson loop, Def. \ref{def:Wilson}}.
 For  $\beta=o_v^2 e o_w^2 \bar e$ ,  the corresponding expectation value \[
\qquad\mathbb E[ \Tr  \hol \beta ]  =
\frac1{\mathcal Z_{\runterhalb{\includegraphics[width=6ex]{quiver_HCIZbez}}\!\!,\,B}  }
\int_{\Dir(B)}  \Tr_{\C^{16}}  \big (\varphi_v^2 U_e \varphi_w^2 U_e^* \big)
\ee^{-16 \cdot \Tr [ q(\varphi_v)  +q( \varphi_w )+ 4 \varphi_v U_e \varphi_w U^*_e ] }
\dif D \nonumber
\]
is an example of a Wilson loop.
\end{enumerate}
The next section verses on how to tackle this kind of integrals
without integration.

\section{The Makeenko-Migdal loop equations for the spectral action} \label{sec:main}

\subsection{Notation} We now derive the constraints on the set of  Wilson loops.
With this aim, we pick an edge
$\de\in Q_1$ which we assume not to be a self-loop, $s(\de) \neq t(\de)$. \par

\begin{figure}
 \includegraphics[width=4.6cm]{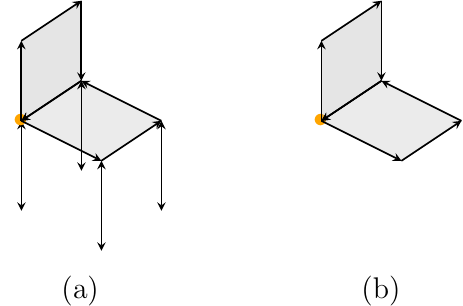}
 \caption{(a) An arbitrary (non-reduced, length-16) loop on a
   rectangular lattice is shown.  The `chair legs' are edges present
   in the combination $e \bar e$.  (b) Its reduced version. Notice
   that both paths have the same holonomy, though the path in (a) is
   larger than (b) in eight (the four removed combinations $e_i\bar
   e_i$ at the legs).  \label{fig:reduced}}
\end{figure}

\begin{figure}%
\includegraphics[width=5.0cm]{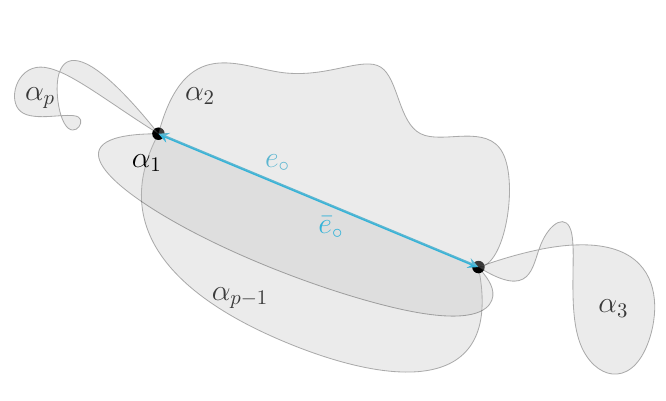}
\caption{The most general reduced path
$\gamma = \deo^{\epsilon_1} \alpha_1 \deo^{\epsilon_2}
   \alpha_2 \cdots  \deo^{\epsilon_m} \alpha_m$ containing $\deo$ and/or $\bde$
\label{fig:GralPathinSA} is shown here (omitting orientations for sake of simplicity).
The most general Wilson loop can also be
decomposed in subpaths in a similar way, cf. eqs. \eqref{Decompose}.}
\end{figure}

 Assume that along a given path $\gamma$ the combinations $e \bar e$
 and $\bar e e$ are \textit{absent} for each edge $e \in \gamma$. We
 call this type of paths \textit{reduced} (Fig. \ref{fig:reduced}) and
 it is trivial to see that reduction of a path (i.e. removing those
 pairs) yields a new one with unaltered holonomy.  Consider then a
 reduced loop $\gamma$ that appears in the spectral action and
 contains the rooted edge $\de$.  This assumption allows (w.l.o.g. due
 to cyclic reordering) the decomposition
\begin{subequations} \label{Decompose}
\[\gamma = \deo^{\epsilon_1} \alpha_1 \deo^{\epsilon_2}
   \alpha_2 \cdots  \deo^{\epsilon_m} \alpha_m =:\textstyle\prod_{i=1}^{m}  \deo^{\epsilon_i} \alpha_i \label{gamma_Decomposition}
  \]
 (cf. Fig. \ref{fig:GralPathinSA}) where each of
  $\epsilon_1,\epsilon_2,\ldots,\epsilon_m \in \{+1,-1\}$ is a sign.
  This convention means that $\deo^{\epsilon}=\bde$ is the edge $\de$
  backwards if $\epsilon=-1$, while of course $\deo^\epsilon$ is $\de$
  itself if $\epsilon=1$.  (The condition that $\gamma$ starts with
  $\de$ implies $\epsilon_1=1$ above, but leaving this implicit is
  convenient.)  By asking that each subpath $\alpha_1,\ldots, \alpha_m
  \subset \gamma$ does not contain neither $\de$ nor $\bde$, one
  uniquely determines the $\alpha_j$'s.  For another loop $\beta$,
  which also starts with $\de$, under the same assumption that $\de$
  and $\bde$ do not appear consecutively in $\beta$ in any order, a
  similar decomposition holds
\[
   \beta &= \deo^{\sigma_1} \mu_1 \deo^{\sigma_2}
   \mu_2 \cdots  \deo^{\sigma_p} \mu_p  \label{beta_Decomposition}
  \]%
\end{subequations}%
in terms of signs $\sigma_j \in \{-1,+1\}$
and paths $\mu_j$ not containing neither the rooted edge $\de$
nor $\bde$. The only difference in notation ---which we will keep throughout---
is that $\gamma$ will refer to generalised plaquettes (i.e. contribution to the spectral action)
while $\beta$ will be the path of a Wilson loop.
\par
Take again the polynomial $f(x)=f_0+ f_1 x + f_2 x^2 + \ldots + f_d x^d $,
and rephrase the spectral action of eq. \eqref{loopexpSA} as
\[ \label{SA}
S(D_Q)= \Tr f(D_Q) =  \sumsub{ \gamma  \in \Omega Q \\ \gamma \text{ reduced }}  g_\gamma  \Tr \hol \gamma.   \qquad
\]
Now $g_\gamma$ is a function of $f_{\ell(\gamma)}$ but possibly also
of $f_{\ell(\gamma) + 2 }, f_{\ell(\gamma) + 4 },\ldots$, whenever
these last coefficients are non-zero. The contribution
of the higher coefficients is owed to the appearance in larger paths of a
contiguous pair of edges $e, \bar e$ for which the respective
unitary matrices will satisfy $U_e U^*_e=1 =U_e^* U_e$.  These
cancellations are not detected by the holonomy, which is the criterion
used in \eqref{SA} to collect all terms (instead of using, as in
eq. \eqref{loopexpSA}, the $f_0,\ldots, f_d$ coefficients and
performing directly the sum over paths).  For instance, if $\gamma$ is
the path in Fig. \ref{fig:reduced} (b), then $g_\gamma$ depends on
$f_{\ell(\gamma)}$ and $f_{\ell(\gamma)+8}$, since
Fig. \ref{fig:reduced} (a) contributes the same to the spectral
action.  The function $g_\gamma=g_\gamma(f_0,f_1,\ldots, f_d)$ is of
course quiver-dependent.

\subsection{Main statement}
The Makeenko-Migdal or loop equations we are about to generalise
appeared first in lattice quantum chromodynamics
\cite{MakeenkoMigdal}. They have been a fundamental ingredient in the
construction of Yang-Mills theory in
\cite{zbMATH06731252, zbMATH06721414, Cao:2023uqm} in rigorous
probabilistic terms.
\\

The main result that follows relates observables of gauge theories on graphs that need not
be lattices. If the next equations  seem abstract, the reader might want to
begin with the fully worked-out case in Section \ref{sec:Applications}. Also Remark
\ref{rem:cases} lists important specific cases of the theorem (and that is why
that remark is before the proof). Then
it will be clear why the next equations are important and why they generalise
those by Makeenko-Migdal.

\begin{theorem}[Makeenko-Migdal equations for the spectral action on quivers]
\label{thm:MM}
 Let $R$ be a representation of a connected quiver $Q$ and let $N=\dim R$.
Root an edge $\de$ of $Q$ that is not a self-loop and abbreviate by
  $U=U_{\de}$ the unitary matrix  that $R$ determines for $\de$. Then
for any reduced loop $\beta $, decomposed as $   \beta  = \deo^{\sigma_1} \mu_1 \deo^{\sigma_2}
   \mu_2 \cdots  \deo^{\sigma_p} \mu_p$  according to eq. \eqref{beta_Decomposition},
   the following relation among Wilson loops holds: \[  \label{MMeqs}
 &\mathbb E \bigg[
\sumsub{j =1  \\ \sigma_j =+1 } ^p  \frac1N\Tr  ( U^{\sigma_1}  \hol \mu_1 \cdots  U^{\sigma_{j-1} } \hol \mu_{j-1} )   \frac1N\Tr   (U^{\sigma_j }  \hol \mu_j\cdots U^{\sigma_p} \hol \mu_p )  \\& \,\, -
 \sumsub{j =1 \\ \sigma_j =-1 }^p   \frac1N \Tr (  \hol \mu_1 U^{\sigma_2}\hol \mu_2\cdots  U^{\sigma_{j-1} } \hol \mu_{j-1}        ) \frac1N \Tr  (\hol \mu_j U^{\sigma_{j+1}}  \cdots U^{\sigma_p} \hol \mu_p )
\bigg]  \notag\\
& = \sumsub{\gamma \in S(D) \\ \gamma \text{ \scriptsize reduced  }
\\ \gamma= \prod_{i=1}^{m(\gamma)}  \deo^{\epsilon_i} \alpha_i
}g_\gamma
\mathbb E \bigg[
\sumsub{j =1  \\ \epsilon_j =+1 } ^{m(\gamma) }
\frac1N
\Tr (
 \hol \beta \cdot U^{\epsilon_j} \hol \alpha_j   \cdots \hol \alpha_m U^{\epsilon_1}  \hol \alpha_1 \cdots U^{\epsilon_{j-1}}\hol\alpha_{j-1} ) \notag
 \\[-2ex]
&\qquad\qquad\qquad\qquad \!-
\sumsub{j =1  \\ \epsilon_j =-1  }  ^{m(\gamma) } \frac1N\Tr ( \hol \beta \cdot
 \hol \alpha_j  U^{\epsilon_{j+1}}  \cdots \hol \alpha_m U^{\epsilon_1} \hol\alpha_1 \cdots \hol\alpha_{j-1}   U^{\epsilon_j} ) \bigg ] \notag,
\]
where the dependence
$\gamma=  \deo^{\epsilon_1(\gamma)} \alpha_1 \deo^{\epsilon_2(\gamma)}
   \alpha_2 \cdots  \deo^{\epsilon_{m(\gamma)}} \alpha_{m(\gamma)} $
on the signs $\epsilon_i$ and the subpaths $\alpha_i$ on $\gamma$
is left implicit  for sake of notation.
\end{theorem}
\begin{remark}\label{rem:cases}Some special cases of eqs. \eqref{MMeqs} are commented on:
 \begin{enumerate}
\item
The second line (lhs) takes the expectation value of
$\tfrac 1N \Tr ( U^{\sigma_1}  \hol \mu_1$ $ \cdots  U^{\sigma_{j-1} }$ $ \hol \mu_{j-1} $ $    U^{\sigma_j }   ) $ $\times \frac1N \Tr  (\hol \mu_j \cdots U^{\sigma_p} \hol \mu_p )$,
but $\sigma_j$ being $-1$ allows for a cancellation, hence the
apparent lack of harmony between the first two lines of the lhs.
\vspace{.51ex}\item
We also stress that the first term in the lhs, which corresponds to
$j=1=\sigma_1$, yields the input Wilson loop $\beta$ in the first
trace and a constant path in the second; the latter yields a factor of
$N$, which is cancelled by its prefactor.
\vspace{.51ex}\item If neither $\bde$ nor $\de$ are along $\gamma$, then
  $m(\gamma)=0$ and the respective sum is empty (the rhs is zero).
\vspace{.51ex}\item Similarly, if neither $\bde$ nor $\de$ are on $\beta$, which is the case
of the constant loop, $p=0$ and the sum in question is empty (the lhs is zero).
\vspace{.51ex}\item
Suppose that the plaquettes in the action $S(D) =
\sum_\gamma^{\text{\tiny reduced}} \tilde g_\gamma [ \Tr \hol \gamma +
  \Tr \hol \bar\gamma] $, intersect each either $\de$ or $\bde$
exactly once. Notice that this time we have rewritten it as sum over
pairs $\gamma$ and $\bar\gamma$ (which is always possible since the
paths are in $\Gamma Q$ and the spectral action is real valued). Then
 \[ \notag&
\mathbb E \bigg[
\sumsub{j =1  \\ \sigma_j =+1 } ^p \frac{1}N \Tr
( U  \hol \mu_1 \cdots  U^{\sigma_{j-1} } \hol \mu_{j-1} ) \frac{1}N  \Tr   (U^{\sigma_j }  \hol \mu_j \cdots U^{\sigma_p} \hol \mu_p )  \\ & \,\,-
 \sumsub{j =1 \\ \sigma_j =-1 }^p \frac{1}N  \Tr (  \hol \mu_1 \cdots  U^{\sigma_{j-1} } \hol \mu_{j-1}        )  \frac{1}N\Tr  (\hol \mu_j \cdots U^{\sigma_p} \hol \mu_p ) \,
\bigg]  \notag\\
  & = \sumsub{\gamma \in S(D) \\ \gamma=(U,\alpha) \text{ \scriptsize reduced  } }\tilde g_\gamma
\mathbb E \bigg[
\frac 1N
\Tr (
 \hol \beta \cdot U  \hol \alpha  )
 -
\frac 1N
 \Tr (
 \hol \beta  \hol \bar\alpha \cdot U^*   ) \bigg] . \label{MMsimplePlaquettes}
\]
 \end{enumerate}

\end{remark}

\begin{proof}
Consider the unitary matrix $U_\de $
associated to the rooted edge
 $\de \in Q_1$,  as given by a fixed
$\pS$-representation $R=\{ (A_v,H_v) ,(\phi_e, U_e)\}_{v,e} $ of
$Q$. Next, consider the infinitesimal variation of the spectral action
by the change of variable exclusively for the unitary matrix $U_\de$ at the
edge $\de$ as follows. Let
\[ \label{transformationU}
U_\de \mapsto U_\de'=  \ee^{\ii Y} U_\de , \qquad  \ii  Y \in \mathfrak{su}(N) ,\qquad \ii=\sqrt{-1},
\]
where $Y$ is given in terms of arbitrary matrices $ \ii y_{k} \in  \mathfrak{su}(n_{t(e),k})$ for $k=1,2,\ldots,
l_{t(\de)}=: L$ by
\[ \notag
 \ee^{\ii Y} := \diag \big[ 1_{r_{t(e),1 } } \otimes \exp({\ii y_1  }),
 1_{r_{t(e),2 } } \otimes \exp({\ii y_2 }), \ldots,
 1_{r_{t(e), { L   } }} \otimes \exp (\ii y_{ L } )  \big ] .
\]
(Recall Sec. \ref{sec:dDandZ} for notation).  One should keep in mind
that this implies also the substitution $U_\de ^* \mapsto (U_\de')^* =
U_\de^* \ee^{-\ii Y}$, as it follows from the change
\eqref{transformationU}.  This rule defines a new representation $R'$
differing from $R$ only by the value of the unitary matrices at the edge
$\de$, that is \[ R'= \{(A_v,H_v) ,(\phi_e, \exp( \delta_{e,\de} \ii Y
) U_{e}\}_{v \in Q_0,e \in Q_1} ,\] where $\delta_{e,e'}$ is the
indicator function on the edge-set. \par

The loop or Dyson-Schwinger or (in the unitary case)  Makeenko-Migdal equations follow from
\[ \label{identity_initial}
\int  \sum_{a,b=1}^N(\partial_Y)_{a,b} \{ (\hol_{R'} \beta)_{b,a} \times \ee^{- N S ( D'  ) } \}
\dif D =0,\qquad D'=D_Q{(R')}.
\]
(The entries of the matrix derivative are $(\partial_Y)_{a,b} =
\partial/\partial Y_{b,a} $ when $Y$ is hermitian.) This follows from the invariance of the
Haar measure at the rooted edge under the transformation
\eqref{transformationU}, yielding $\dif D'=\dif D$. Below, we show
that this implies
\[
\mathbb E \bigg[  \Big( \frac1N \Tr\otimes \frac 1N \Tr \Big) (\partial_Y  \hol \beta ) \bigg]
=\mathbb E \bigg[ \frac 1N \Tr (\partial_Y S \hol \beta ) \bigg],
\]
and compute each quantity inside the trace(s). On the lhs, the matrix
derivative acts on a noncommutative polynomial and is then the
Rota-Stein-Turnbull noncommutative derivation\footnote{ This means, in
terms of entries, $ \partial /\partial {Y_{b,a}} (Y^k)_{r,s} =
\sum_{l=0}^{k-1} (Y ^k\otimes Y^{k-1-l})_{r,a| b,s} = \sum_{l=0}^{k-1}
Y^l_{r,a} Y^{k-1-l}_{b,s} $ writing out the noncommutative derivative.
}
\[ \label{ncder}
\partial_Y  Y^{k+1}  =\sum_{l=0}^{k}    Y^l \otimes Y^{k-l} \qquad (Y^*=Y \in \M N, k \in\Z_{\geq 0})
\]
while on the rhs, the matrix yields Voiculescu's cyclic derivation
$\mathcal D_Y$, since the quantity it derives, $S$, contains a trace.
Such derivative $\mathcal D_{Y_j}$ is defined, say for $q\in \N$, on
the free algebra $\C_{\langle q \rangle} =\C \langle
Y_1,\ldots,Y_q\rangle $ on a monic noncommutative monomial $\psi$ by
\[ \mathcal D_{Y_j} \psi(Y_1,\ldots,Y_q) = \sumsub{P,\Lambda \in \C_{\langle q \rangle} \\ \psi = \Lambda Y_j  P } P \Lambda. \label{Voiculescuder} \]
(The sum is performed over all splittings by $Y_j$ of the word $\psi$
\cite[Sec. 7.2.2]{Guionnet}, wherein $P$ or $\Lambda$ might be empty). \par
Recalling that holonomies are multiplicative, one has $ \hol \gamma =
U ^{\epsilon_1} \hol \alpha_1 U^{\epsilon_2} \hol \alpha_2$ $
U^{\epsilon_3} $ $ \cdots \hol \alpha_{m-1}$ $ U ^{\epsilon_m} \hol
\alpha_m$.  With respect to the transformed representation $R'$ we can
compute the holonomy $\hol_{R'} \delta$ of any path $\delta$. This
depends on $Y$ and $\de$ but we use a prime in favor of a light
notation and write $\hol' \delta$.  Since none of the subpaths
$\alpha_j$ contains the transformed edges $\de$ and $\bde$, one has
$\hol' \alpha_j = \hol \alpha_j$, so
 \[\hol' (\gamma)
 =U ^{'\epsilon_1}
 \hol  \alpha_1
 U^{'\epsilon_2}
 \hol  \alpha_2 U^{'\epsilon_3} \cdots
  U ^{'\epsilon_m} \hol  \alpha_m
 \]
 where $U^{'\epsilon }$ is $\ee^{\ii Y} U $ if $\epsilon=1$
 and $U^*\ee^{-\ii Y}$ if $\epsilon=-1$.
Therefore the variation of the loop $\gamma$ writes
\[ \notag
[\partial_Y \Tr \hol' \gamma ] \big |_{Y=0}
& =
 \ii \sumsub{j =1 \\ \epsilon_j = +1 } ^m
U  \hol  \alpha_j U^{\epsilon_{j+1}}
\hol  \alpha_{j+1} \cdots
U^{\epsilon_n} \hol  \alpha_n
 U^{\epsilon_{1} } \hol  \alpha_1 \cdots
U^{\epsilon_{j-1}} \hol  \alpha_{j-1} \\ &
-
\ii
\sumsub{j =1 \\ \epsilon_j = -1 }^m
 \hol  \alpha_j U^{\epsilon_{j+1}}
\hol  \alpha_{j+1} \cdots
U^{\epsilon_n} \hol  \alpha_n
 U^{\epsilon_{1} } \hol  \alpha_1 \cdots
U^{\epsilon_{j-1}} \hol  \alpha_{j-1} U^* \notag.
\]
The cyclic wandering of any fix holonomy, say $\hol \alpha_1$, in the rhs of
the main result is due to Voiculescu's cyclic derivation \eqref{Voiculescuder}.
\par
We now compute the variation of the Wilson line $\beta$, whose
holonomy writes for the representation $R'$ as
$
\hol' \beta =\prod_{j=1}^p (U')^{\sigma_j} \hol' \mu_j
=
\prod_{j=1}^p (U')^{\sigma_j} \hol \mu_j $.
 To take the variation observe that $\hol'\beta$ is not inside a trace.
For any $A,B\in \M N$, and $a,b,c,d=1,\ldots, N$, due to eq. \eqref{ncder},
\[
(\partial_Y)_{a,b}\nonumber
[ A \exp(   \ii Y ) B ]_{c,d} \bigg|_{Y=0}
& = \sum_{k=0}^\infty \frac{\ii^k}{k!}
\sum_{l=0}^k   A_{c,r}  [ Y^l \otimes Y^{k-1-l}] _{ r,a |b,s }\bigg|_{Y=0}  B_{s,d} \\
  & =\ii A_{c,r} \delta_{r,a} \delta_{b,s}\nonumber
B_{s,d}.
\]
Using this rule for the previous expression of  $
\hol' \beta $, one obtains a summand for each occurrence of $U^{\pm1 }$
and the result follows after equating the indices $c=a$, and $b=d$, which
is the initial situation in the initial identity \eqref{identity_initial}.
\end{proof}
%


\subsection{Graphical representation of the Makeenko-Migdal equations}
We illustrate graphically the meaning of the Makeenko-Migdal equations.
Let us place $\de$ and the reversed edge $\bde$ along a fixed axis of the picture.
To represent a  Wilson loop $\beta$ or a reduced
generalised plaquette $\gamma$, we choose the following notation.
In order to avoid drawings with several intersections,
for each time that $\gamma$ or $\beta$ walks along either $\de$ or $\bde$,
we jump to the next `plane' in anti-clockwise direction around the fixed axis.
Thus each of these planes
represents abstractly the subpath $\mu_j \subset \beta$
or $\alpha_j\subset \gamma$ according to the decompositions
 \eqref{gamma_Decomposition}
and  \eqref{beta_Decomposition}, that is:
\[ \notag
\includegraphics[width=.54\textwidth]{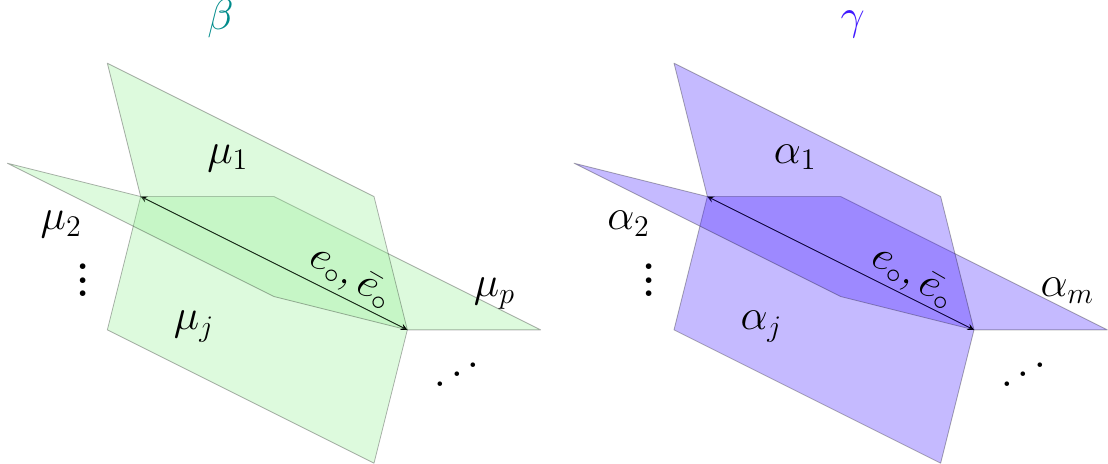}
\]
We kept a rectangular appearance for sake of visual simplicity, but
the subpaths $\mu_j$ and $\alpha_j$ are arbitrary (as far as they have
positive length).  In fact, the depicted situation is due to a second
reason still oversimplified: the theorem describes the more general
case that $\mu_j$ or $\alpha_j$ might be loops themselves (as
$\alpha_3$ in Fig. \ref{fig:GralPathinSA}), but this would render the
pictures unreadable. The representation of the Makeenko-Migdal
equations reads then as follows:

\[
\text{ $\frac1{N^2} \times \mathbb E$}
\Vast(\raisebox{-.5\height}{
\includegraphics[width=.33998\textwidth]{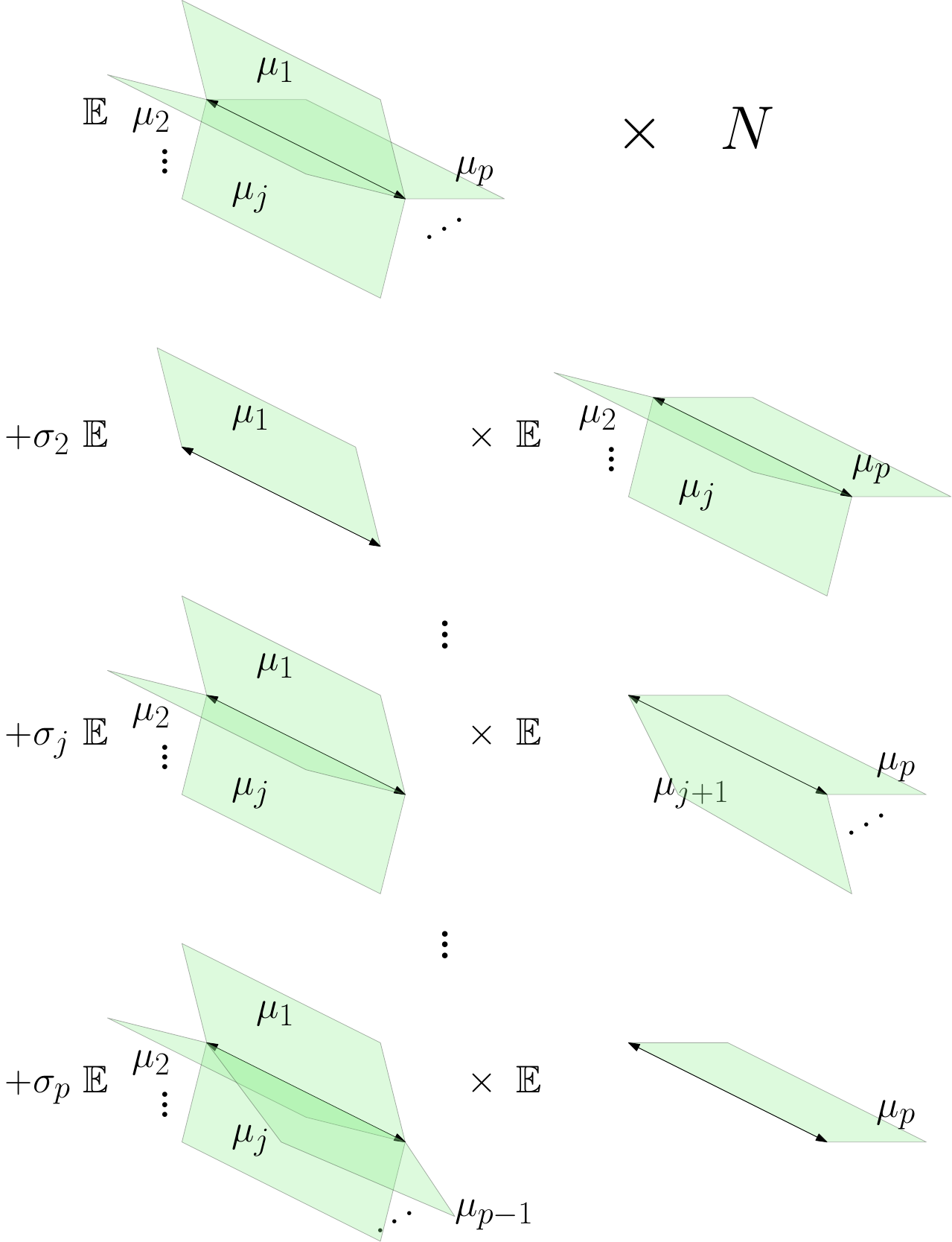}}\Vast)\,
\text{  $=  \sum_\gamma  g_\gamma  \mathbb E$  }
\vast( \!\!\!
\raisebox{-.5\height}{
\includegraphics[width=.303\textwidth]{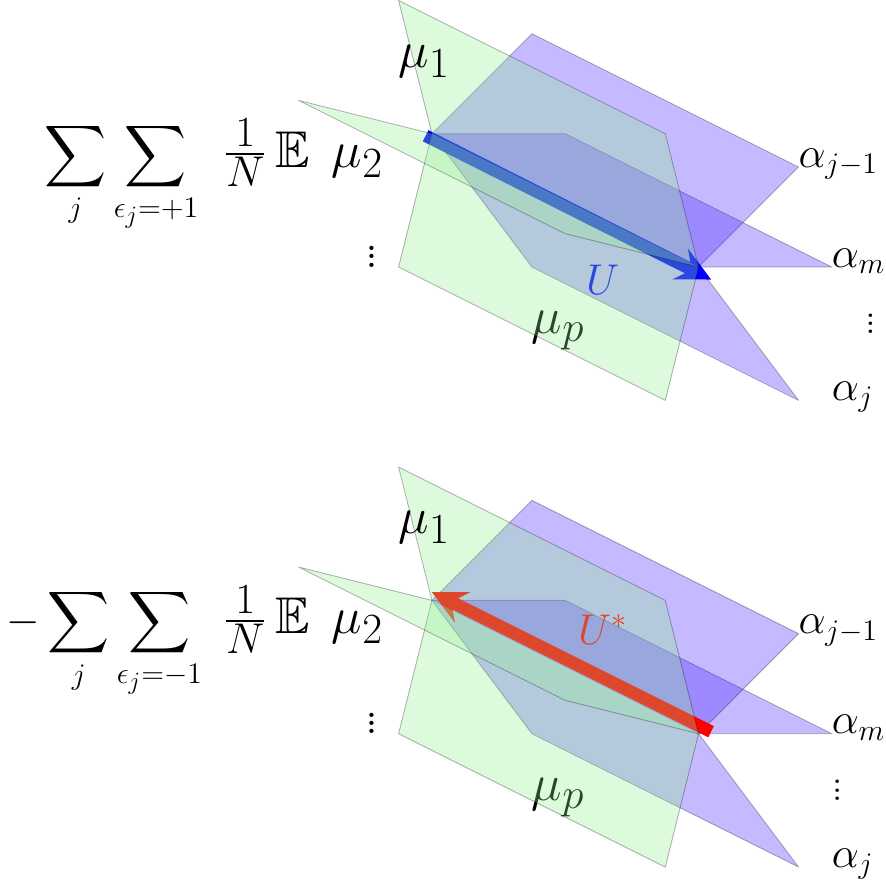}}\vast) \notag
\]
In the rhs, the very similar upper and lower terms need a word of
notation.  The blue arrow denotes an insertion of $U$ and is executed
right after the green part of the path, while the red arrow inserts
$U^*$ and follows only after the purple set of paths.

\section{Applications} \label{sec:Applications}
This last section aims at illustrating the power of the equations
derived here when combined with the positivity conditions.  This
combination, sometimes known as `bootstrap', appeared in
\cite{Kruczenski} for lattice gauge theory and \cite{LinBootstrap} in
a string context (for hermitian multimatrix models).

\subsection{Positivity constraints}
\begin{figure}
\includegraphics[width=7cm]{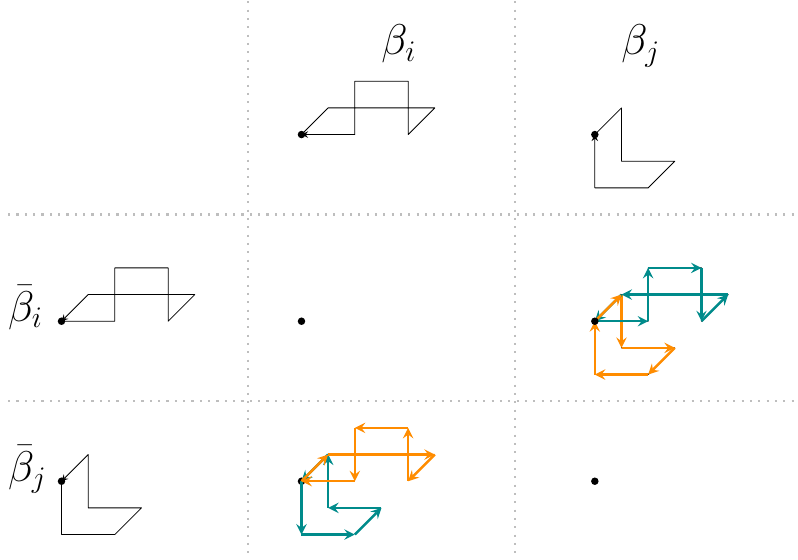}
\caption{The matrix entries for a $2\times 2$ submatrix of $\mathcal
  M$ (before taking expectation values) are shown in this figure.  On
  the diagonal the trivial entries one sees result from the (holonomy
  of both) loops cancelling, which has the holonomy of the constant
  path.  On the off-diagonal entries, one sees a non-trivial loop
  composition that goes first around the orange arrows and then along
  the green ones.
\label{fig:LoopMatrix}}
\end{figure}

Let $v\in Q_0$ be fixed for this subsection and fix a representation $R$ of $Q$
of dimension $N$.
Consider a complex variable $z_\beta$ for each
loop $\beta$ based at a fixed vertex $v \in Q_0$, $ z=\{ z_\beta :  \beta \in \Omega_v( Q) \}$,
as well as the matrix
\[
P( z ) :=  \sumsub{\beta \in \Omega_v(Q)} z_\beta \hol \beta, \qquad P( z ) \in \M N.
\]
It follows that $
\Tr  \big[ P(z)  P(z)^* \big] =
\sum_{\beta, \alpha } z_\beta z^*_\alpha
\hol \beta \cdot  (\hol \alpha)^*
=
\sum_{\beta, \alpha\in \Omega_v(Q) }
z_\beta z^*_\alpha
\hol( \beta  \bar \alpha) \geq 0$
independently of the $z$-tuple; this is preserved by
expectation values, i.e.
\[
\sum_{\beta, \alpha\in \Omega_v(Q) }
z_\beta z^*_\alpha
\mathbb E[ \hol( \beta  \bar \alpha) ]  \geq 0, \qquad  \text{for all } z \in \C^{\Omega_v(Q)},
\]
which is an equivalent way to state the positivity $\mathcal M \succeq  0$ of the matrix
$\mathcal M \in \C [\![N,f_0,f_1,\ldots,f_d ]\!]$
whose entries are given by
\[ \label{Mmatrix}
(\mathcal M )_{i,j} := \mathbb E[ \hol( \beta_i  \bar \beta_j ) ]
\]
for any ordering of the loops $\{\beta_1,\beta_2,\ldots\} \subset
\Omega_v(Q)$ at the fixed vertex $v$, as illustrated by
Figure \ref{fig:LoopMatrix}. The positivity of $\mathcal M$
is clearly independent of the way we order these loops, as a
conjugation by a permutation matrix (which is a unitary
transformation) will not change the eigenvalues of $\mathcal M$.  \par

The paths $\beta_i$ and $\beta_j$ feeding the matrix \eqref{Mmatrix}
need only to satisfy $s(\beta_i)=s(\beta_j)$ and
$t(\beta_j)=t(\beta_i)$ so that $\beta_i\bar \beta_j$ is a loop;
the assumption that $\beta_i $  and $\beta_j$ themselves are loops is not essential. The
choice for the matrix \eqref{Mmatrix} with loop entries is originally
from \cite{Kazakov:2024ool}, who pushed forward the bootstrap for
lattice Yang-Mills theory.  The techniques of \cite{LinBootstrap} were
implemented for fuzzy spectral triples for an interesting kind of
hermitian matrix \cite{Hessam:2021byc} and a hermitian 2-matrix model
\cite{Khalkhali:2023onm}.  The loop equations of \cite{MakeenkoMigdal}
have been extended here to include arbitrary plaquettes that whirl
around any edge more than once, and Wilson loops that are allowed to
do the same. \par
\subsection{A complete example}

Consider the triangle quiver $
Q=\raisebox{-.81ex}{\includegraphics[height=1.02cm]{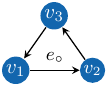}}
$ with a rooted edge $\de$, and let $ \zeta = \de \mu$ be the only
loop of length 3 starting with $\de$ ($\mu$ is the path $v_2\to v_3\to
v_1$, of course). Fix the the Bratteli network $B$ given by
$A_{v_i}=\M N, H_{v_i}=\C^N$ for the three vertices, $i=1,2,3$ (the
transition matrices $C_e$ have all one entry equal to 1, for the three
vertices).  The space $\Dir(B)$ of Dirac operators is therefore three
copies of $\uni(N)$, and the corresponding partition function
\[\mathcal Z_{Q,B}= \int_{\uni(N)^3}
\ee^{-N S(D(U_1,U_2,U_3)) }
\dif U_1\dif U_2 \dif U_3 .  \label{ZDreieck}\]
The action $S(D) = \Tr f(D)$ for
$f(t)=f_0 + f _1  t+f_2 t^2+ f_3 t^3$ with real coefficients reads
\[
S(D) =(f_0+2 f_2)N +
x [
\Tr \hol\zeta + \Tr \hol \zeta\inv ],
\]
where we set $x =3 f_3$.  The terms in the even coefficients are just
constants that disappear when evaluating Wilson loops; we therefore
set $f_0=f_2=0$.
\subsubsection{Loop equations}
Now pick a loop $\beta= \zeta^n$ for positive $ n\in
\Z$. According to the loop equations \eqref{MMsimplePlaquettes}, one
has
\[
\mathbb E \bigg[ \sum_{k=0}^{n-1}  (\frac 1N \Tr \otimes \frac 1N \Tr) (\hol \zeta^k \otimes \hol \zeta^{n-k})\bigg]
= \frac xN (\mathbb E \Tr \hol \zeta^{n+1} -
\mathbb E \Tr\hol  \zeta^{n-1} ) .
\]
Defining the large-$N$ moments by $m_j:= \lim_{N\to \infty} \mathbb E [ \tfrac 1N\Tr \hol  \zeta^{j}]$ for each $j\in \Z$,
this means
\[
\sum_{l=0}^{n-1}
m_l  \cdot m_{n-l}
=
x (m_{n+1}
-m_{n-1}  ) , \qquad (N\to \infty),
\]
since large-$N$ factorisation holds, $N^{-2} \mathbb E[ \Tr\hol  \zeta^i \Tr\hol  \zeta^j]  \to
m_i \cdot m_j
$, as $N\to \infty$.
For the loop $\beta= \zeta^{-n}$
with $n\in\N$, one has
\[ \label{negativemoments}
-\sum_{j=0}^{n-1}
m_{-(n-j)} \cdot m_{-j}
=x (  m_{-(n-1)}  -  m_{-(n+1)}),  \qquad (N\to \infty).
\]
Finally, going through the derivation of the loop equations for the
constant Wilson loop, one obtains the vanishing of the lhs, so $0= x
(m_1 - m_{-1})$, hence $\bar m_1 = \overline{\mathbb E [ \Tr \hol
    \zeta ] } = {\mathbb E [ \Tr \hol \bar \zeta ] } = {\mathbb E [
    \Tr \hol \zeta \inv ] }=m_{-1} = m_1$, so $m_1$ is real (this can
be derived by other means, but the loop equations yield this
explicitly).  Together with eq. \eqref{negativemoments}, this implies
$m_{-j} = m_j$ for all $j=1,2,\ldots$ and the moments can be arranged
in the following (due to $\mathcal
M_{i,j}=\mathcal M_{i+k,j+k}$, Toeplitz-)matrix:
\[
\mathcal M =
\begin{bmatrix}
 1 & m_1 & m_2 & m_3 & \ldots & \\
 m_{-1} &  1 & m_1 & m_2  & \ldots & \\
m_{-2} & m_{-1} &
 1 & m_1 &   \ldots & \\
m_{-3} & m_{-2} & m_{-1} &
 1 &    \ldots & \\
\vdots & \vdots & \vdots & \vdots & \ddots
 \end{bmatrix}=
\begin{bmatrix}
 1 & m_1 & m_2 & m_3 & \ldots & \\
 m_1 &  1 & m_1 & m_2  & \ldots & \\
m_2 & m_ 1 &
 1 & m_1 &   \ldots & \\
m_3 & m_2 & m_ 1 &
 1 &    \ldots & \\
\vdots & \vdots & \vdots & \vdots & \ddots
 \end{bmatrix}.
\]
\subsubsection{Bootstrap}
Thanks to Theorem \ref{thm:MM}, $\mathcal M$ can be computed
recursively in terms of $y:=m_1$ and the coupling $x$,
\[ \notag
 m_1 & = y &&& m_4& = \frac{4  y}{x} + \frac{3  y^{2}}{x^{2}} + \frac{1}{x^{2}} + \frac{y}{x^{3}} + 1 \\ m_2 &  = \frac{y}{x} + 1
 &&&  \notag
  m_5&  = y + \frac{3  y^{2}}{x} + \frac{2  y^{3}}{x^{2}} + \frac{3}{x} + \frac{9  y}{x^{2}} + \frac{6  y^{2}}{x^{3}} + \frac{1}{x^{3}} + \frac{y}{x^{4}}
 \\ m_3& = y + \frac{y^{2}}{x} + \frac{1}{x} + \frac{y}{x^{2}}
  &&& m_6 & =\frac{9  y}{x} + \frac{18  y^{2}}{x^{2}} + \frac{10  y^{3}}{x^{3}} + \frac{6}{x^{2}} + \frac{16  y}{x^{3}} + \frac{10  y^{2}}{x^{4}} + \frac{1}{x^{4}} + \frac{y}{x^{5}} + 1 . \notag
\]

\begin{figure}\centering
\includegraphics[width=.99\textwidth]{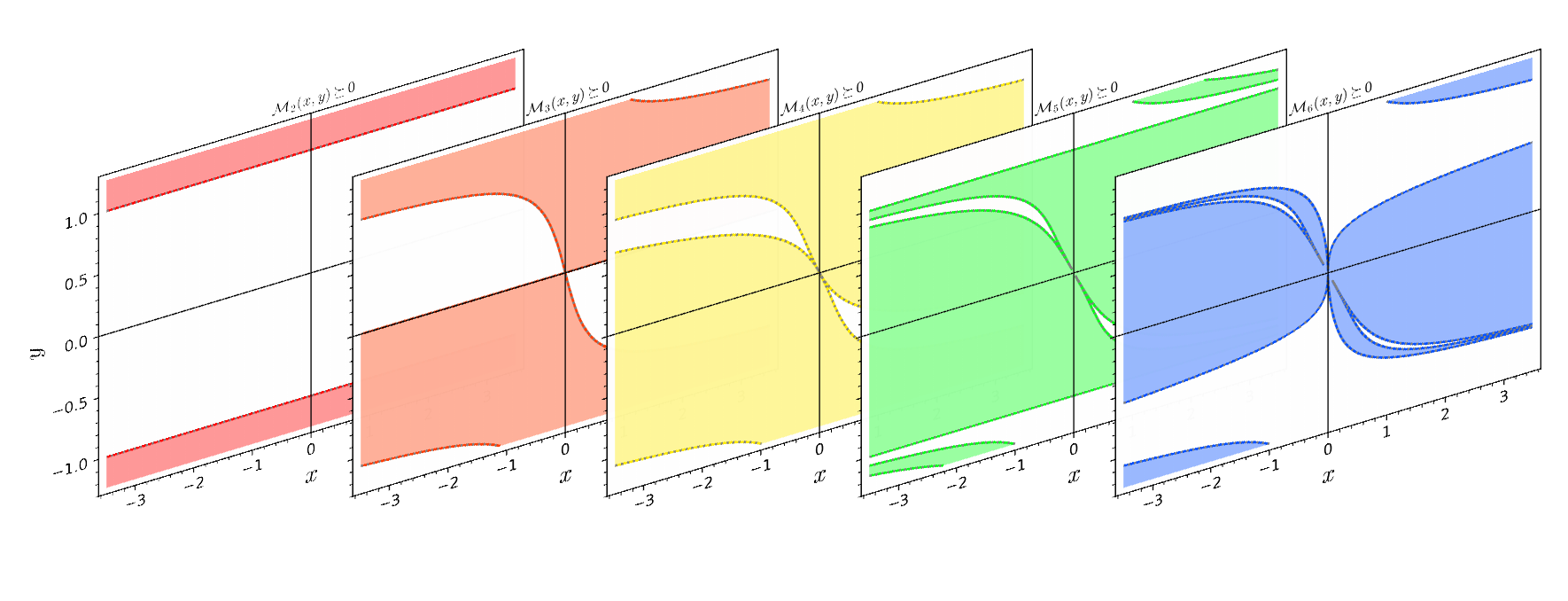}\vspace{-2ex}
 \includegraphics[width=.58\textwidth]{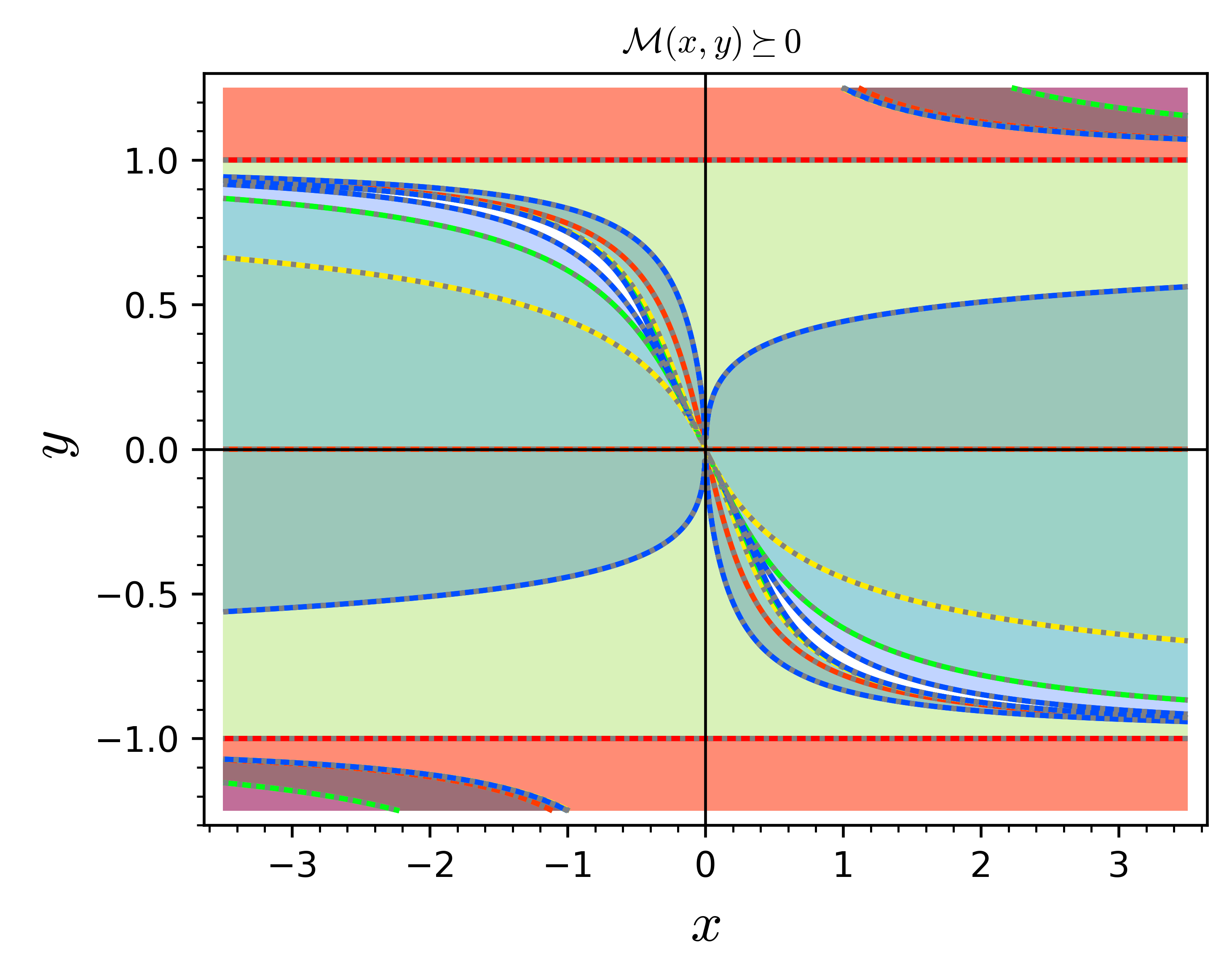}
 $\qquad \quad $
 \caption{\label{fig:minors1to6} On the panel above, the colored
   regions violate $ \mathcal M (x,y)\succeq 0$ as a consequence of
   the minors $\mathcal M_n (x,y) $ satisfying $\det \mathcal M_n
   (x,y) < 0$ for $n=1$ (tautological, not drawn), $n=2$ (the
   complement to the stripe $|y |<1$), $n=3$ (orange), $n=4$ (yellow),
   $n=5$ (green) and $n=6$ (blue).  Superposition of all these plots
   yields the plot below, in which only in the narrow white region
   satisfies the simultaneous conditions $\det \mathcal M_n (x,y)> 0$,
   $n=1,2,\ldots, 6$. Fig. \ref{fig:comparison} goes further, but due
   to readability shows only the seventh minor, which narrows down
   even more the white space. Script and plots use \texttt{SageMath} \cite{sage}.\vspace{-2ex}}
\end{figure}

The positivity condition $ \mathcal M(x,y) \succeq 0$ can be plotted
on the \textit{first moment} vs. \textit{coupling} plane in terms of
the simultaneous positivity of its minors $\mathcal M_n(x,y):=
[\mathcal M(x,y)_{a,b}]_{a,b=1,\ldots,n}$.  as done in Figure
\ref{fig:minors1to6} for $n=1,\ldots, 6$.
\subsubsection{Exact solution}
Let us contrast this strategy with the analytic solution. The
partition function \eqref{ZDreieck} can be simplified by
integrating\footnote{\label{Razvan}The author thanks R\u azvan Gur\u au for this
remark.} over a single unitary group, with the change of variable $U= U_1U_2U_3$,
  \[ \nonumber
 \mathcal Z_{Q,B} & =
 \int _{\uni(N)^3}
 \ee^{-Nx \Tr ( U_1U_2U_3 +U_3^*U_2^*U_1^*)}  \dif U_1 \dif U_2 \dif U_3 \\
 &=\Bigg( \int _{\uni(N) }\dif U_1 \Bigg)\Bigg(\int _{\uni(N)}  \dif U_2 \Bigg)\int_{\uni(N)} \nonumber
 \ee^{-Nx \Tr (U  +U ^*)} \dif U \\ &= \int _{\uni(N) }
 \ee^{-Nx \Tr (U +U^*)} \dif U=:\mathcal Z_N(x). \label{defZ}
 \]
We now contrast the positivity constraints with the exact solution by
Wadia and Grosse-Witten (GWW).  Their strategy was to diagonalise the
integration variable as $U= V \Theta V^*$, by a $V\in \uni (N)$, being
$\Theta = \diag(\ee^{\ii \theta_1},\ee^{\ii \theta_2},\ldots, \ee^{\ii
  \theta_N}) \in \uni (1)^N$. This yields an integral over the torus
$\uni (1)^N$ of $\prod_{j=1}^N\ee^{-2Nx \cos \theta_j} \times
\prod_{1\leq i<k \leq N} |\ee^{\ii \theta_i }-\ee^{\ii \theta_k}|
^2$. The last factor is of the form $\det (\Delta)\det(\Delta^*)$,
where $\Delta_{m,n}=\exp( n \ii \theta_m) $ is the Vandermonde matrix
from the change of variable. The explicit expression solution is \cite{Wadia,GrossWitten}
 \[
 \mathcal Z_N(x)
 &=
\det [I_{k-m} (-2xN)  ]_{k,m=1,\ldots, N} \\ & :  = \det \begin{pmatrix}
                                      I_0(z) & I_1(z) & \cdots & I_{N-1} (z) \\
                                       I_{-1}(z) & I_0(z) & \cdots & I_{N-2}  (z) \\
                                          \vdots &  \vdots & \ddots & \vdots  \\
                                            I_{-(N-1)}(z) & I_{-(N-2)}(z) & \cdots & I_{0}(z)
                                     \end{pmatrix} \nonumber
_{z=-2xN}
 \]
   for the partition function as the determinant of a Toeplitz matrix
 of Bessel $I$-functions,
 \[ I_{q}(z) :=
\frac{1}{2\pi}\int_0^{2\pi}
\ee^{\ii q \alpha + z\cos \alpha }
\dif \alpha, \]
evaluated at $z=-2xN$. Armed with this explicit solution, the
exact moment $y_N= \mathbb E [ \frac 1N \Tr  \hol \zeta ] $
by eq. \eqref{defZ} reads (the expectation values of $\Tr \hol \zeta$ and $\Tr \hol \zeta\inv$
 coincide, hence the factor $\tfrac12$)
\[ y_N(x)= -\frac{1}{2 \mathcal Z_N(x) N^2} \frac \partial{ \partial x} \mathcal Z_N(x)  .\]
\begin{figure}[h!]\centering
 \includegraphics[width=.69\textwidth]{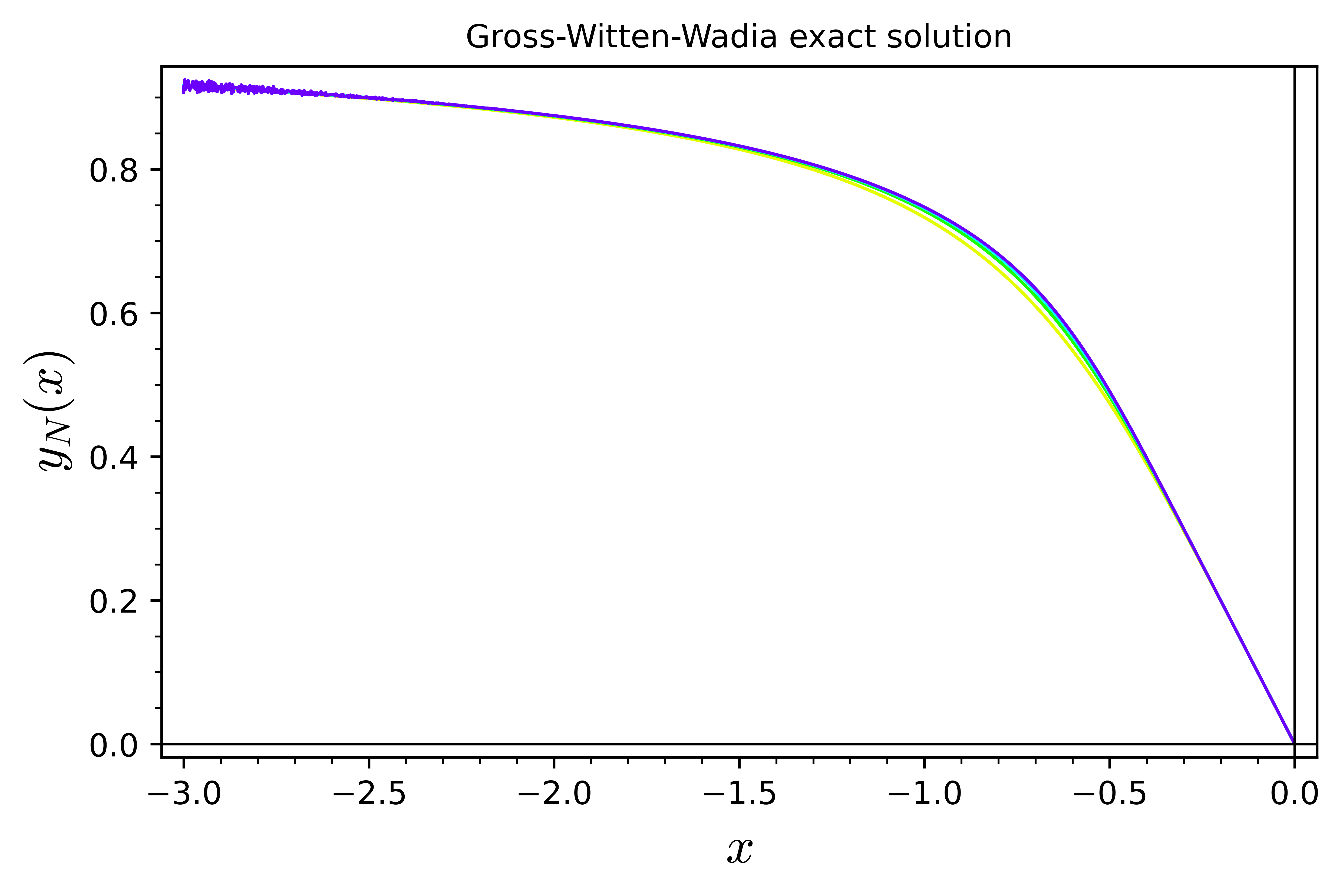}
 \caption{The expectation value of $\frac1N \Tr \hol \zeta= \frac1N \Tr
   [U_1U_2U_3 ]$ computed from the exact partition function $ \mathcal
   Z_N(x)$ via $y_N(x)= (-1/2 \mathcal Z_N(x) N^2) \partial_x \mathcal
   Z_N(x) $ at finite $N$, namely for $ N \in \{
   \color{YellowGreen!50!yellow}2,\color{green!60!cyan}3,
   \color{cyan}4, \color{RoyalPurple} 5\color{black}\}$.
   Cf. comparison with the bootstrap solution in
   Fig. \ref{fig:comparison}.
 \label{fig:exact}
 }
\end{figure}%

\begin{figure}[h!]\centering
 \includegraphics[width=.73\textwidth]{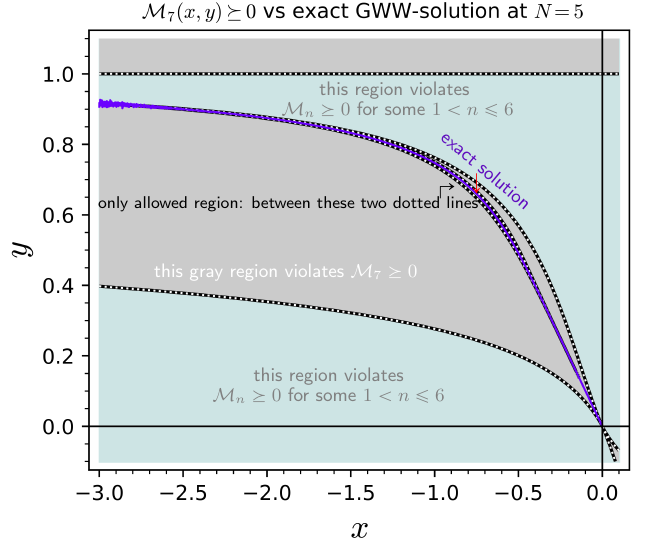}\qquad
 \qquad
 \caption{\label{fig:comparison} Comparison of the darker gray region
   that violates $\mathcal M_7(x,y)\succeq 0 $ with the exact
   GWW-solution at $N=5$ in solid color line. The large light blue
   regions around the gray bulk are excluded by testing the minors
   $\mathcal M_n(x,y)$ for $n\leq 6$. This leaves only a narrow
   allowed region between the `parallel' paths tagged with an arrow;
   there sits the exact solution, plotted here for $N=5$. (Observe however
      [coming from the negative direction towards the origin]  that about \rojo{at} $x=-0.15$ the parallel lines inside the gray region do close before they
   arrive to the origin. This seems to be an artifact plot-software.)}
\end{figure}

This was plotted for different values of $N$ in Figure
\ref{fig:exact}.  If our loop equations are correct, then the curve $
y_N(x)$ should lie inside the region where $\mathcal M(x,y)$ is
non-negative for large enough $N$ (agreement only at large $N$ is
expected since freeness or factorisation of the expectation values was
used to compute the matrix of moments and $\mathcal M$).  Luckily, this
is what clearly happens in the plots of Figure \ref{fig:comparison}:
the highest technically feasible computation for $\mathcal M(x,y)$
yielded a very tight constraint where the expectation value $y_N$
computed from the GWW partition function embeds.

\subsection{Concluding remarks and outlook}\label{sec:conclusion_outlook}
The results of this article can be summarised as follows.
Given the two first elements of the spectral triple $A_Q,H_Q$
associated to a quiver $Q$ (equivalent to a Bratteli network on $Q$),
we characterised the ensemble of Dirac operators $D$
that complete $(A_Q,H_Q,D)$ into a spectral triple, as well as the
measure $\dif D$ on such ensemble. The partition function
\vspace{.2ex}
\[
\mathcal Z_{A,H} (f) = \int_{\!\!\!\!\substack{\phantom{nic tu} \\[1ex] (A,H,D) \text{ \scriptsize is} \\ \text{ \scriptsize spectral triple}}} \ee^{- N  S(D)}\dif D,
\quad S(D)=\Tr f(D), \quad N= \dim H.
  \]
is made concrete here.
Since $\dif D$ is a Haar measure, unitary invariance leads to
constraints for the Wilson loops of this theory.
Such loop equations were proven and applied in combination with positivity conditions
in the case of  a simple example.
\\

As happened above, the observed situation for a large class of
hermitian matrix integrals are tight constraints for the first moment
(or for a finite set of moments) in terms of the coupling, which, by
increasing the size of the minors, typically determine a curve
$y=y(x)$---and by the respective loop equations, all the moments and
thus the solution of the model.  In this article we do not claim the
convergence of a `bootstrapped' solution in all ensembles of unitary
matrices. The aim of this example was to illustrate the usefulness of
the loop equations proven here.  But the results of this example do
encourage us to explore this combination in future works, including
also a hermitian (`Higgs scalar') field that arises from the
self-loops of the quiver. \\

When a family of random quivers is chosen to
discretise a particular space, another type of convergence
that is crucial to consider is that of the quiver themselves, namely
in a generalised sense of Lov\'asz-Szegedy \cite{Lovasz} convergence
to graphons or limit graphs\footnote{The author thanks
an anonymous referee for the references\label{ungarische_autoren}.}. Thereby
it is essential to take care of
the discretisation's artifacts
(e.g. there exist parasitic solutions emerging in the discretised
version of the ODE to
Euler's 1744 buckling problem; such parasites are  `kinky',
approximate no smooth solution, and can be taken care of by methods
explained in \cite[Sec. 4]{Domokos}).


\vspace{4ex}
{ \footnotesize
\section*{Acknowledgements} \footnotesize
 This work was mainly supported by the European Research
Council (ERC) under the European Union’s Horizon 2020 research and
innovation program (grant agreement No818066) and also by the Deutsche
Forschungsgemeinschaft (DFG, German Research Foundation) under
Germany’s Excellence Strategy EXC-2181/1-390900948 (the Heidelberg
\textsc{Structures} Cluster of Excellence).  I thank the Erwin
Schrödinger International Institute for Mathematics and Physics (ESI)
Vienna, where this article was finished, for optimal working
conditions and hospitality.  I acknowledge the kind answers by the
group of Masoud Khalkhali at Western U., specially by Nathan
Pagliaroli, on a question about bootstrapping. I thank Thomas
Krajewski and R\u azvan Gur\u au for valuable comments, specially the latter for motivating the
comparison with the exact solution. {The author also acknowledges
insightful comments by three referees (see footnotes \ref{fn:Ref2}, \ref{ungarische_autoren}).}}

\section*{Data availability statement}
Data generated in this article are available upon reasonable request.


\bibliographystyle{alpha}

\end{document}